\documentclass{article}

\usepackage{arxiv}

\usepackage[utf8]{inputenc} 
\usepackage[T1]{fontenc}    
\usepackage{hyperref}       
\usepackage{url}            
\usepackage{booktabs}       
\usepackage{amsfonts}       
\usepackage{nicefrac}       
\usepackage{microtype}      
\usepackage{lipsum}
\usepackage{graphicx}
\graphicspath{ {./images/} }
\usepackage{authblk}
\usepackage{caption}
\usepackage{subcaption}
\usepackage{multirow}
\usepackage{rotating}
\usepackage[acronym,nonumberlist,style=super,toc]{glossaries}
\usepackage{glossary-mcols}

\makenoidxglossaries

\newacronym{AI}{AI}{Artificial Intelligence}
\newacronym{ML}{ML}{Machine Learning}
\newacronym{DL}{DL}{Deep Learning}

\newacronym{ACC}{ACC}{Accuracy}
\newacronym{GM score}{GM score}{Geometric Mean Score}
\newacronym{PREC}{PREC}{Precision}
\newacronym{REC}{REC}{Recall}
\newacronym{AUC}{AUC}{Area Under Curve}
\newacronym{TP}{TP}{True Positives}
\newacronym{TN}{TN}{True Negatives}
\newacronym{FP}{FP}{False Positives}
\newacronym{FN}{FN}{False Negatives}
\newacronym{TPR}{TPR}{True Positive Rate (Sensitivity)}
\newacronym{TNR}{TNR}{True Negative Rate (Specificity)}
\newacronym{FPR}{FPR}{False Positive Rate}
\newacronym{FNR}{FNR}{False Negative Rate}
\newacronym{ROC}{ROC}{Receiver Operating Characteristic}

\newacronym{KS score}{KS score}{Kolmogorov-Smirnov score}
\newacronym{MSE}{MSE}{Mean Squared Error}
\newacronym{CPH}{CPH}{Cox Proportional-Hazards Model}
\newacronym{RSF}{RSF}{Random Survival Forest}
\newacronym{MTLR}{MTLR}{Multi-Task Logistic Regression}
\newacronym{N-MTLR}{N-MTLR}{Nested Multi-Task Logistic Regression}
\newacronym{DeepSurv}{DeepSurv}{Deep Survival Analysis}
\newacronym{NB}{NB}{Naive Bayes}
\newacronym{GBT}{GBT}{Gradient Boosting Tree}
\newacronym{Boosted DT}{Boosted DT}{Boosted Decision Tree}
\newacronym{PR curve}{PR curve}{Precision-Recall curve}
\newacronym{CSF}{CSF}{Conditional Survival Forest}
\newacronym{SHAP}{SHAP}{SHapley Additive exPlanations}
\newacronym{XAI}{XAI}{Explainable Artificial Intelligence}

\newacronym{GBM}{GBM}{Gradient Boosting Machine}
\newacronym{NN}{NN}{Neural Network}
\newacronym{RNN}{RNN}{Recurrent Neural Network}
\newacronym{GRU}{GRU}{Gated Recurrent Unit}
\newacronym{MLP}{MLP}{Multilayer Perceptron}
\newacronym{XGBoost}{XGBoost}{Extreme Gradient Boosting}
\newacronym{LightGBM}{LightGBM}{Light Gradient Boosting Machine}
\newacronym{CatBoost}{CatBoost}{Categorical Boosting}

\newacronym{Ref}{Ref}{Reference}
\newacronym{LR}{LR}{Logistic Regression}
\newacronym{DT}{DT}{Decision Tree}
\newacronym{RF}{RF}{Random Forest}
\newacronym{AdaBoost}{AdaBoost}{Adaptive Boosting}
\newacronym{KS}{KS}{Kolmogorov-Smirnov}
\newacronym{TDL}{TDL}{Total Deviation from Linearity}
\newacronym{KNN}{KNN}{K-Nearest Neighbors}
\newacronym{SVM}{SVM}{Support Vector Machine}
\newacronym{ANN}{ANN}{Artificial Neural Network}
\newacronym{LSTM}{LSTM}{Long Short-Term Memory}
\newacronym{IOHMM}{IOHMM}{Input-Output Hidden Markov Model}
\newacronym{GNB}{GNB}{Gaussian Naive Bayes}
\newacronym{CNN}{CNN}{Convolutional Neural Network}
\newacronym{Weibull}{Weibull}{Weibull Distribution}
\newacronym{Gompertz}{Gompertz}{Gompertz Distribution}
\newacronym{C-index}{C-index}{Concordance Index}
\newacronym{IBS}{IBS}{Integrated Brier Score}
\newacronym{MAE}{MAE}{Mean Absolute Error}
\newacronym{DNN}{DNN}{Deep Neural Network}
\newacronym{AUCPR}{AUCPR}{Area Under the Precision-Recall Curve}
\newacronym{SMOTE}{SMOTE}{Synthetic Minority Over-sampling Technique}

\usepackage{hyperref}
\usepackage{color}
\hypersetup{%
colorlinks=true,
linkcolor=blue,
citecolor=blue,
urlcolor=blue}

\usepackage{hyperref}
\usepackage{bookmark}
\usepackage{etoolbox}
\makeatletter

\usepackage{float}
\restylefloat{table}

\def\tsc#1{\csdef{#1}{\textsc{\lowercase{#1}}\xspace}}
\tsc{WGM}
\tsc{QE}
\tsc{EP}
\tsc{PMS}
\tsc{BEC}
\tsc{DE}

\title{AI-based identification and support of at-risk students: A case study of the Moroccan education system}

\author{Ismail ELBOUKNIFY\textsuperscript{1}\thanks{ismail.elbouknify@um6p.ma}, Ismail BERRADA\textsuperscript{1}, Loubna MEKOUAR\textsuperscript{1}, Youssef IRAQI\textsuperscript{1}, EL Houcine BERGOU\textsuperscript{1}, Hind BELHABIB\textsuperscript{2}, Younes NAIL\textsuperscript{2}, Souhail WARDI\textsuperscript{2}}

\affil{\textsuperscript{1} College of Computing, Mohammed VI Polytechnic University, Benguerir, Morocco\\
\textsuperscript{2} Ministry of National Education, Preschool, and Sports, Rabat, Morocco}

\begin{document}
\maketitle
\begin{abstract}
Student dropout is a global issue influenced by personal, familial, and academic factors, with varying rates across countries. This paper introduces an AI-driven predictive modeling approach to identify students at risk of dropping out using advanced machine learning techniques. The goal is to enable timely interventions and improve educational outcomes. Our methodology is adaptable across different educational systems and levels. By employing a rigorous evaluation framework, we assess model performance and use Shapley Additive exPlanations (SHAP) to identify key factors influencing predictions. The approach was tested on real data provided by the Moroccan Ministry of National Education, achieving 88\% accuracy, 88\% recall, 86\% precision, and an AUC of 87\%. These results highlight the effectiveness of the AI models in identifying at-risk students. The framework is adaptable, incorporating historical data for both short and long-term detection, offering a comprehensive solution to the persistent challenge of student dropout.
\end{abstract}

\section{Introduction}

\begin{figure}[htbp]
  \centering
  \includegraphics[width=0.6\textwidth]{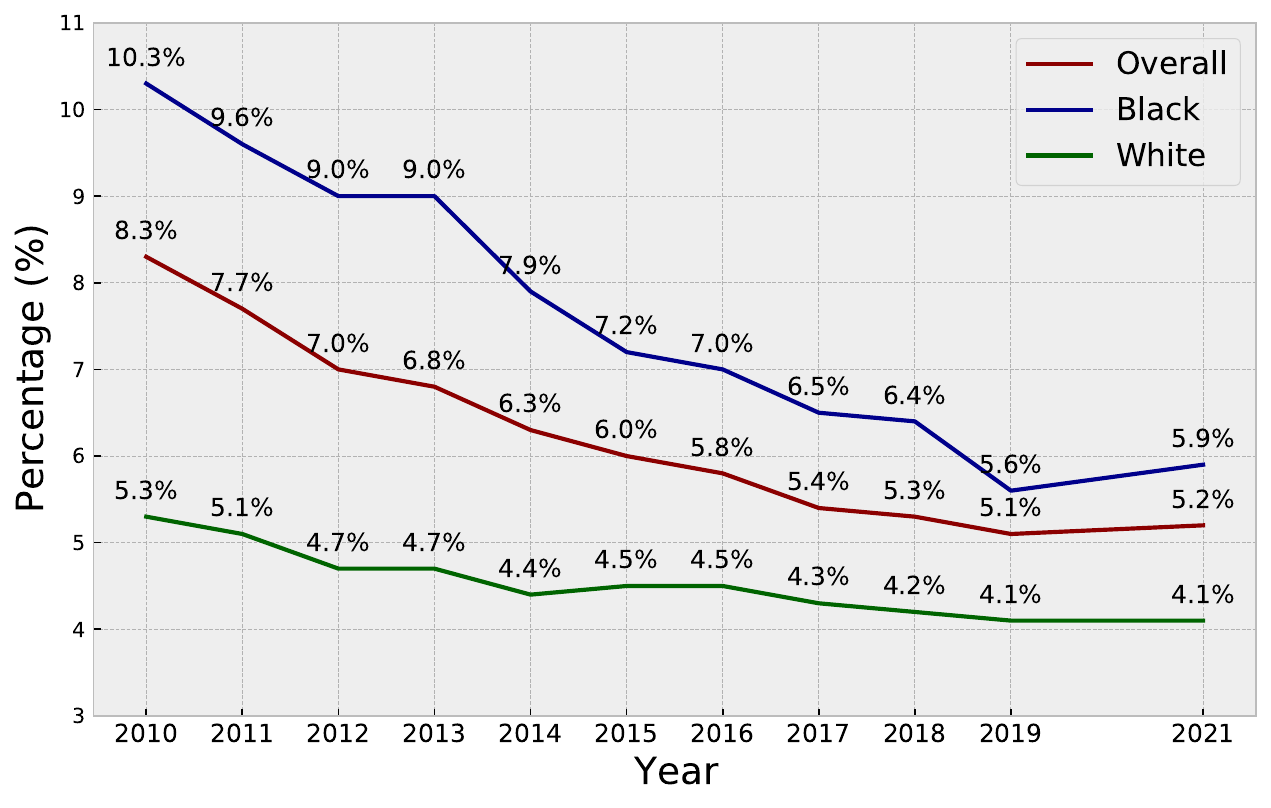}
  \caption{Dropout rates for 16-to-24 year olds by race in the USA}
  \label{fig:Statistics}
\end{figure}

Education is universally acknowledged as a fundamental right and a vital tool for personal growth, social mobility, and economic progress. However, the persistent issue of student dropout poses a significant barrier to achieving these objectives \cite{pusztai2022factors}. Educational institutions worldwide grapple with alarmingly high dropout rates, leading to far-reaching consequences for individuals, families, communities, and societies at large. 
Countries are making concerted efforts to improve educational outcomes \cite{srivani2024design} and provide comprehensive support to ensure that students complete their studies successfully \cite{sepahvand2023adaptive}. A notable example is the United States, which has shown a consistent downward trend in dropout rates, as shown in Figure \ref{fig:Statistics}\footnote{\url{https://nces.ed.gov/}}.
Comprehending the underlying reasons for student dropout is essential to devise effective strategies and interventions \cite{rodriguez2023methodology}. This knowledge empowers educators, policymakers, and stakeholders to develop targeted initiatives aimed at reducing dropout rates and cultivating an inclusive and equitable education system, with repercussions extending beyond individual educational attainment. High dropout rates perpetuate a cycle of limited opportunities and socioeconomic disparities, hindering gainful employment and impacting families, communities, and society as a whole \cite{aina2022determinants}. Moreover, addressing this issue is not only of paramount importance but also presents a unique opportunity in the era of data-driven decision-making, allowing us to leverage advanced analytics and machine learning techniques for a deeper understanding and the development of proactive intervention strategies \cite{pusztai2022factors}. Thus, unraveling the complexities of student dropout is academically stimulating and holds substantial real-world implications, making it a critical and engaging area of research and inquiry.

Addressing the pervasive challenge of student dropout has spurred innovative approaches in education. While some researchers employ traditional data analysis techniques to understand the reasons behind dropout \cite{dol2023classification}, others harness the potential of machine learning and deep learning methods \cite{adelman2018predicting}. By integrating machine learning into education to enhance educational outcomes \cite{pal2012mining}, this approach involves leveraging machine learning techniques to create personalized support systems and targeted interventions, utilizing extensive student data to detect early warning signs of potential dropouts \cite{khan2021random}. This proactive strategy empowers educators and counselors to intervene promptly with tailored support, addressing underlying issues and mitigating dropout risks \cite{rodriguez2023methodology} \cite{ameri2016survival}. Extensive research has already dissected the factors contributing to student dropout, highlighting the influence of variables such as the level of study and the learning environment \cite{hung2019improving}. These multifaceted factors necessitate a holistic approach to fostering an inclusive and supportive educational environment that effectively mitigates dropout rates \cite{gutierrez2023supporting}. Some researchers are also delving into machine learning and deep learning methods to enhance their understanding and prediction of dropout trends \cite{adelman2018predicting}, \cite{fu2021clsa}. In this collective effort to combat student dropout, educational stakeholders combine human expertise with advanced analytical techniques. By doing so, they collaboratively tackle the issue, creating an environment that ensures every student has the opportunity to succeed in their educational journey \cite{kruger2023explainable}. This fusion of traditional and cutting-edge approaches exemplifies the commitment to comprehensively address the complex problem of student dropout.


This paper presents a versatile framework for modeling predictive systems on school dropouts, offering a novel and adaptable solution to the pressing issue of student dropout. With a strong emphasis on versatility, this framework is designed to accommodate a wide range of educational systems, regardless of their unique characteristics and challenges. The proposed framework comprises three main components: 
    \begin{itemize}
      \item \textbf{(1) Data Preprocessing:} In this stage, the data will undergo cleaning, and various feature engineering techniques will be applied.
      \item \textbf{(2) Prediction:} Machine learning models and techniques for handling imbalanced data will be employed to address this issue. Additionally, a prediction corrector is proposed to enhance the precision of the dropout class predictions.
      \item \textbf{(3) Intervention:} This phase will involve interventions based on predictive analysis and practical experience.
    \end{itemize}

Equally important is the framework's ability to seamlessly integrate historical data from different time periods. This feature not only enhances the comprehensiveness of the proposed approach but also enables a dynamic understanding of dropout trends and patterns over time. To rigorously evaluate the efficacy of our proposed solution, we conduct comprehensive testing within a real-world educational system. This testing involves the practical application of our framework to a dataset provided by the Moroccan Ministry of National Education, Preschool, and Sports. By doing so, we aim to assess the real-world applicability and effectiveness of our solution in addressing the critical issue of student dropout. In this study, we have introduced an innovative and scalable approach that holds the potential for widespread applicability across various educational systems. This approach not only represents innovation but also emphasizes its scalability, thus ensuring its effectiveness across a diverse range of educational systems. By testing our proposed framework within the context of the Moroccan education system, we have established its potential for transferability to other educational systems similar to the Moroccan education system, like the United Arab Emirate \cite{gallagher2019education}, Saudi Arabia, France \footnote{\url{https://www.scholaro.com/db/countries/france/education-system}}, Chile \cite{rodriguez2023methodology}.

Our research unveils key insights into student dropout, encompassing temporal patterns, model performance, prediction correction, prediction horizon, and the pivotal role of historical data selection.

\begin{itemize}
  \item Temporal patterns of dropout: Our analysis reveals that the highest rates of dropout consistently coincide with the final stages of each educational cycle.
\item Impact of dropout rates on model performance: We observe a direct correlation between dropout rates and model performance, largely due to the challenges of imbalanced datasets. It is worth noting that when we have larger samples of dropout cases, the models become more adept at detecting the patterns associated with dropout.
  \item Predictive correction for enhanced precision: The integration of our proposed prediction correction mechanism demonstrates a substantial improvement in the precision of identifying students at risk of dropout.
  \item Precision enhancement with extended prediction horizons: The precision of identifying potential dropout cases increases as we extend the prediction horizon.
  \item Significance of historical data selection: The choice of the number of years of historical data significantly influences the performance of our dropout prediction model, highlighting the importance of thoughtful historical data selection.

  \end{itemize}

This paper introduces a novel predictive modeling approach to address student dropout. Our contributions include:
\begin{itemize}
  \item A general framework for identifying at-risk students.
  \item Versatile applicability across diverse educational systems.
  \item Rigorous evaluation through multiple test plans.
  \item A prediction corrector to enhance dropout classification.
  \item Validation using a proprietary dataset from the Moroccan Ministry of National Education, Preschool, and Sports. 
\end{itemize}

These contributions are significant advancements in addressing the complex and pressing issue of student dropout, offering a versatile, rigorously evaluated, and practically applicable solution that can revolutionize educational outcomes.


This paper is organized as follows: Section \ref{sec:Related Work} delves into related work on dropout prediction using AI, presenting the existing literature and approaches in this domain. Section \ref{sec:Framework} presents the general framework and its composition. In Section \ref{sec:UseCase}, we provide an overview of the Moroccan education system and also detail the dataset used in our study, including its characteristics and sources. In Section \ref{sec:Experiments}, we present the results of our experiments and analyze their implications. Section \ref{sec:Discussion}, we engage in a comprehensive discussion of our results, their significance, and potential avenues for future research. Finally, section \ref{sec:Conclusion} offers the conclusions drawn from our research findings.

\section{Related Work}\label{sec:Related Work}
This section reviews key research areas in student dropout prediction. It covers the use of artificial intelligence to identify at-risk students, the diverse input features adopted, the range of predictive models, from traditional machine learning to deep learning, and the evaluation metrics used to assess performance. Moreover, strategies for handling imbalanced data. Additionally, it highlights the varying temporal scopes of historical data usage, framing the current landscape and challenges in dropout prediction research.

\subsection{ Dropout identification based on \gls{AI}}
\gls{AI} is being used in several areas of education, including the analysis of student performance \cite{rabin2019modeling,gomez2024understanding}, early identification of students at risk \cite{rodriguez2023methodology}. The inclusion of \gls{AI} in the realm of education holds great promise for addressing numerous challenges and amplifying learning outcomes \cite{pal2012mining, fan2023predicting}. A particularly significant area of emphasis revolves around combatting the issue of student dropout \cite{rodriguez2023methodology}. \gls{AI}-centered interventions in education seek to harness the potential of data analytics, predictive modeling, and personalized learning strategies. The aim is to provide timely and precise support to students who are susceptible to dropping out \cite{singh2022predicting}. By meticulously scrutinizing an extensive array of academic and non-academic indicators \cite{rodriguez2023methodology} \gls{AI} systems can uncover patterns and cues that signify potential risks of dropout \cite{kruger2023explainable}. This proactive identification empowers educators and administrators to intervene promptly, providing customized measures to mitigate these risks.

\subsection{Adopted Input features}
In the realm of predictive modeling to address student dropout, the selection of input features plays a pivotal role in determining the accuracy and effectiveness of the models \cite{nagy2018predicting,li2022multi}. Different studies have adopted diverse categories of features to capture the multidimensional aspects contributing to student attrition \cite{naseem2019using}, \cite{ai2020dropout}. These input features shed light on a range of factors encompassing demographics, socioeconomic circumstances, academic performance, motivation, and more.
Several research papers \cite{tan2015prediction}, \cite{balta2022perceived}, \cite{costa2017evaluating}, \cite{migueis2018early}, \cite{phan2023decision}, \cite{kumar2022evaluate}, \cite{guner2014predicting} have integrated demographic information such as age, gender, and ethnicity, along with socioeconomic factors like family income and parental education level. These features provide insights into the background and context within which students are pursuing their education. The academic journey of a student \cite{balta2022perceived}, \cite{costa2017evaluating}, \cite{phan2023decision}, \cite{xing2019dropout} is often marked by various indicators,  including past grades, attendance records, and performance in assessments. Additionally, the motivation to engage and succeed in educational pursuits is a crucial aspect that certain studies have taken into account \cite{adelman2018predicting}, recognizing its influence on dropout tendencies. The educational institution itself can influence student retention. Hence, certain studies consider institutional features like class size, available resources, and teaching methodologies \cite{pereira2017application}.
The variation in the choice of input features across different studies underscores the complexity of the dropout phenomenon \cite{zafari2021practical}. The amalgamation of these diverse features in predictive models fosters a multidimensional approach, enhancing the models' ability to accurately identify students at risk of dropping out \cite{moreno2020temporal}. Understanding the categories of features employed offers valuable insights into the depth and comprehensiveness of these predictive approaches.

\subsection{Prediction Models}
The prediction methods employed to identify students at risk of dropout encompass a diverse array of techniques rooted in the realm of \gls{AI} \cite{singh2022predicting}. These methods leverage the power of data-driven insights and machine learning algorithms to proactively recognize potential dropout cases. Several studies have contributed to the development and application of these techniques \cite{chung2019dropout}, shedding light on their effectiveness in enhancing student retention rates \cite{kruger2023explainable}.
Researchers \cite{xing2019dropout} have harnessed a spectrum of machine learning algorithms, such as \gls{DT} \cite{kotsiantis2013decision}, \gls{SVM} \cite{cortes1995support}, \gls{LR} \cite{hosmer2013applied}, to predict at-risk students. These algorithms delve into historical academic and contextual data to decipher patterns and indicators that suggest potential dropouts. The work by \cite{ortigosa2019lab} is noteworthy in this regard, employing a \gls{RF} \cite{breiman2001random} classifier to accurately identify students at risk of dropout based on a comprehensive set of input features.
The advent of \gls{DL} \cite{lecun2015deep} has ushered in a new era of predictive modeling in education \cite{olive2019quest}. \gls{DNN}, including \gls{CNN} \cite{gutierrez2023supporting}  and \gls{RNN} \cite{fei2015temporal}, have been applied to capture intricate relationships within data. The study conducted by \cite{prenkaj2021hidden} stands out, employing a \gls{DL} architecture to analyze sequential data and predict student attrition with high accuracy.
These AI-driven techniques serve as valuable tools for institutions seeking to implement proactive measures to retain students \cite{hung2019improving}. By identifying at-risk students and providing timely interventions, these methods contribute to enhancing the overall educational experience and bolstering student success \cite{smith2020global}.

\subsection{Evaluation Metrics}
Various metrics are employed to comprehensively assess how well the models perform in accurately categorizing students based on their likelihood of dropping out. It is imperative to carefully choose metrics that align with the specific goals and nuances of the educational context. Many studies predominantly \cite{tan2015prediction}, \cite{manrique2019analysis}, \cite{chung2019dropout}, employ metrics such as Accuracy, Precision, Recall, and F1-score  while other research papers \cite{rodriguez2023methodology}, \cite{lee2019machine}, \cite{queiroga2020learning} explore alternative metrics like \gls{AUC}, Sensitivity, and Specificity. Interestingly, certain studies in the field acknowledge the limitations of traditional metrics, such as accuracy, when dealing with imbalanced datasets. For instance, \cite{rodriguez2023methodology} emphasizes the need to calculate metrics by class in imbalanced data scenarios. The consideration of class-specific metrics, particularly in imbalanced data scenarios \cite{gaudreault2021analysis}, offers a more accurate depiction of a model's performance, ensuring its relevance and effectiveness.

\subsection{Imbalanced Classification Problems}

In the context of identifying students at risk of dropout, the challenge of imbalanced data distribution poses a significant hurdle \cite{manrique2019analysis}. The prevalence of a majority class compared to a minority class can lead to biased model training and reduced performance in capturing the patterns of interest \cite{chung2019dropout}. To counter this challenge, various techniques have been employed to enable the models to make accurate predictions across both classes \cite{liu2023imbalanced}. 
To mitigate the imbalanced data challenge, resampling techniques are employed in several papers, focusing on either oversampling \cite{chung2019dropout} the minority class or undersampling \cite{nagrecha2017mooc} the majority class. Some of these papers use the oversampling technique \cite{sharma2022review}. This technique involves creating synthetic instances of the minority class to balance the class distribution. The most widely utilized technique for this purpose is \gls{SMOTE} \cite{lee2019machine} which generates synthetic instances by interpolating between existing instances, addressing the data scarcity issue for the minority class \cite{chawla2002smote}. Other papers use the Undersampling technique  \cite{du2020integrated}, \cite{manrique2019analysis}. This technique entails reducing the number of instances in the majority class to balance the distribution \cite{mohammed2020machine}. This prevents the model from being biased towards the majority class and encourages it to learn from both classes more equitably. 

In other studies, the technique of class weighting, as described in \cite{cano2013weighted}, was used. This involves assigning varying weights to classes during the model training process, with higher weights allocated to the minority class. This weight assignment strategy ensures that the model places greater emphasis on accurately classifying instances belonging to the minority class \cite{aljohani2023novel}.
Ensemble models like the random forest, \gls{XGBoost}, \gls{LightGBM}, and \gls{CatBoost} are also employed to tackle the issue of imbalanced datasets, as discussed in \cite{rodriguez2023methodology}, \cite{song2023all}. Ensembling involves aggregating predictions from multiple models to enhance overall predictive performance \cite{coleman2019better}. These ensemble methods achieve this by training each model on distinct resampled datasets, leveraging the diversity of these models to make more precise predictions.
Addressing the imbalanced data challenge is pivotal to building effective predictive models for student dropout prediction \cite{werner2023imbalanced}. Resampling techniques, class weighting, ensembling, and hybrid approaches are key tools in rebalancing the data distribution and enhancing model accuracy. By applying these techniques and evaluating models with class-specific metrics, researchers can effectively navigate the complexities of imbalanced data in the pursuit of accurate dropout prediction.

\begin{sidewaystable*}[htp]
 \centering
    \caption{Summary of recent important works in predicting students at risk of dropout}
    \label{tab:works_summary}
    \resizebox{1\hsize}{!}{$
    \begin{tabular}{cccccccc}
    \toprule
    
    Ref & Target Level & Features category  & Dataset size  & Models & Metrics & Imbalanced data  \\
    &&&&&&techniques\\
    \midrule
    
    \cite{rodriguez2023methodology} & Chilean education system & Academic, demographic & 3 million & \gls{DT}, \gls{XGBoost}, \gls{LightGBM}, & Precision, Recall, & Not specified \\
    &&& & \gls{CatBoost} & F1-score, \gls{GM score}\\
    
    \cite{kruger2023explainable} &  Preschool, Primary school & Academic, demographic &  29972 & \gls{DT}, \gls{LR}, \gls{RF}, \gls{AdaBoost},  & \gls{AUC}, Precision, Recall, & Usage of metrics \\
    & Secondary school &  sosio-economic &  & \gls{XGBoost} & \gls{KS score} &\\

    \cite{migueis2018early} & High school & Academic, demographic  & 2459 & \gls{RF}, \gls{DT} & Precision, Recall & Not specified \\

    \cite{phan2023decision} & Bachelor, master & Demographic, academic,  & 14391 & \gls{LR} & \gls{AUC}, \gls{TDL} & Not specified \\
     & &  student feedback &   \\

    \cite{xing2019dropout} & Online courses  & Academic, platform data & 3617 & \gls{KNN}, \gls{SVM}, \gls{DT}, & \gls{AUC}, Accuracy & Not specified \\
    &&&& \gls{MLP} \\
    
    \cite{chung2019dropout} & High school & Academic, demographic & 165715 & \gls{LR}, \gls{SVM}, \gls{ANN}, \gls{RF} & Confusion Matrix & Not specified \\

    \cite{ortigosa2019lab} & Online courses & Demographic, & 11000 & \gls{RF}, \gls{DT} & Recall, \gls{FPR}, \gls{AUC} & Not specified \\
     && platform data, economic & \\

    \cite{fei2015temporal} & Online courses & Academic, platform data & 39877 & \gls{LSTM},   & \gls{AUC} & Not specified \\
    &&&& \gls{IOHMM},\\
    &&&& \gls{SVM}, \gls{LR}, \gls{RNN}\\
     
    \cite{gutierrez2023supporting} & University & Academic, demographic,  & 13696 & \gls{LR}, \gls{SVM}, \gls{GNB}, \gls{KNN},   & Accuracy, \gls{AUC}, \gls{MSE},  & Not specified \\
     &&  socio-economic &  & \gls{DT}, \gls{RF}, \gls{CNN}, \gls{Weibull},  & \gls{C-index},  \\
     &&&&& \gls{IBS},\\
     &&&&&  \gls{MSE}, \gls{MAE}\\
     
     &&&& \gls{Gompertz},\\
     &&&& \gls{CPH},\\
     &&&& \gls{RSF},\\
     &&&& \gls{CSF},\\
     &&&&  \gls{MTLR}, \\
     &&&&  \gls{N-MTLR},\\
     &&&&  \gls{DeepSurv}\\

     \cite{olive2019quest} & Online courses & Platform data & 78722
     &  \gls{DNN} & Accuracy, F1-score & Not specified \\

     \cite{prenkaj2021hidden} & Online courses & Platform data & 249000
     &   \gls{CNN}, \gls{LSTM}, \gls{DNN}, \gls{LR},  & \gls{AUCPR}, & Not specified \\
      & & &  & \gls{DT}, \gls{GNB}, \gls{KNN}, \gls{RF}, \gls{SVM} &  F1-score\\

     \cite{hung2019improving} & Online courses &  Platform data & 13368 & \gls{SVM}, \gls{DNN}, \gls{RF}, \gls{DT} & Accuracy, & Not specified \\
      &&&&& Sensitivity\\

     \cite{manrique2019analysis} & University & Academic & 2175 & \gls{NB}, \gls{SVM}, \gls{RF},  & Accuracy, Precision,  & Not specified \\
     &&&  & \gls{GBT}, \gls{KNN} & Recall, F1-score\\
     
     \cite{lee2019machine} & High school & Academic & 16571 & \gls{RF}, \gls{Boosted DT} & \gls{AUC}, \gls{PR curve}  & \gls{SMOTE} \\
     
     \cite{nagrecha2017mooc} & Online  courses & Platform data  & 16000 & \gls{LR}, \gls{DT}, \gls{RF}, \gls{GBT} & \gls{AUC} & Undersampling \\

    \cite{du2020integrated} & Online courses & Platform data  & 11688 & \gls{DT}, \gls{DNN}, \gls{AdaBoost} & Accuracy, F1-score & \gls{SMOTE},  \\
    &&&&&& Undersampling\\

    \cite{song2023all} & University &Academic, demographic  & 60010 & \gls{XGBoost}, \gls{LightGBM}, \gls{DT}, \gls{LR}, \gls{RF} & Accuracy, Precision, & \gls{SMOTE} \\
     &&  socio-economic &  && Recall, F1-score \\
    \bottomrule
    \end{tabular} $}
\end{sidewaystable*}

\subsection{Historical data usage}

In the realm of predicting student dropouts, researchers have adopted diverse timeframes, including week-based \cite{al2019detecting}, \cite{yin2020power}, \cite{wang2021mooc},\cite{poellhuber2023cluster}, \cite{ding2019transfer}, semester-based \cite{tan2015prediction}, \cite{manrique2019analysis}, \cite{chen2018running}, \cite{ameri2016survival}, \cite{song2023all}, \cite{phan2023decision}, \cite{berka2021bachelor}, \cite{guzman2022implementation}, \cite{ortigosa2019lab}, and year-based strategies \cite{rodriguez2023methodology}, \cite{orooji2019predicting}, \cite{ortigosa2019lab}, \cite{singh2022predicting}. These temporal perspectives offer valuable insights into dropout patterns at distinct phases of a student's academic journey. While short-term prediction models have gained prominence for their capacity to promptly identify immediate dropout risks, it is imperative to acknowledge that dropout risks may persist beyond the short term. Certain students may confront challenges that could lead to dropouts in subsequent years. Therefore, a holistic dropout prevention strategy should encompass both short-term interventions for immediate risks and long-term support mechanisms to address enduring challenges. 
To summarize, Table \ref{tab:works_summary} provides a concise overview of some significant related research. Diverse methodologies are introduced, targeting educational systems and online courses, and employing a wide array of feature categories, including academic, demographic, socio-economic, and behavioral attributes. These methodologies primarily leverage \gls{ML} and \gls{DL} models, with diverse performance metrics to assess model effectiveness. Addressing imbalanced data, some papers incorporate techniques like \gls{SMOTE} and undersampling, whereas others do not explicitly address this challenge. Evaluations often rely on random data splitting or k-fold cross-validation, which may not comprehensively reflect model performance, especially for imbalanced educational data. Given the need for practical deployment and adaptability to new datasets for a new academic year, it is crucial to consider news-splitting strategies.


\section{Global Framework Overview}\label{sec:Framework}

\subsection{Student Status Tracking}

\begin{figure}[htb!]
  \centering
  \includegraphics[width=0.6\textwidth]{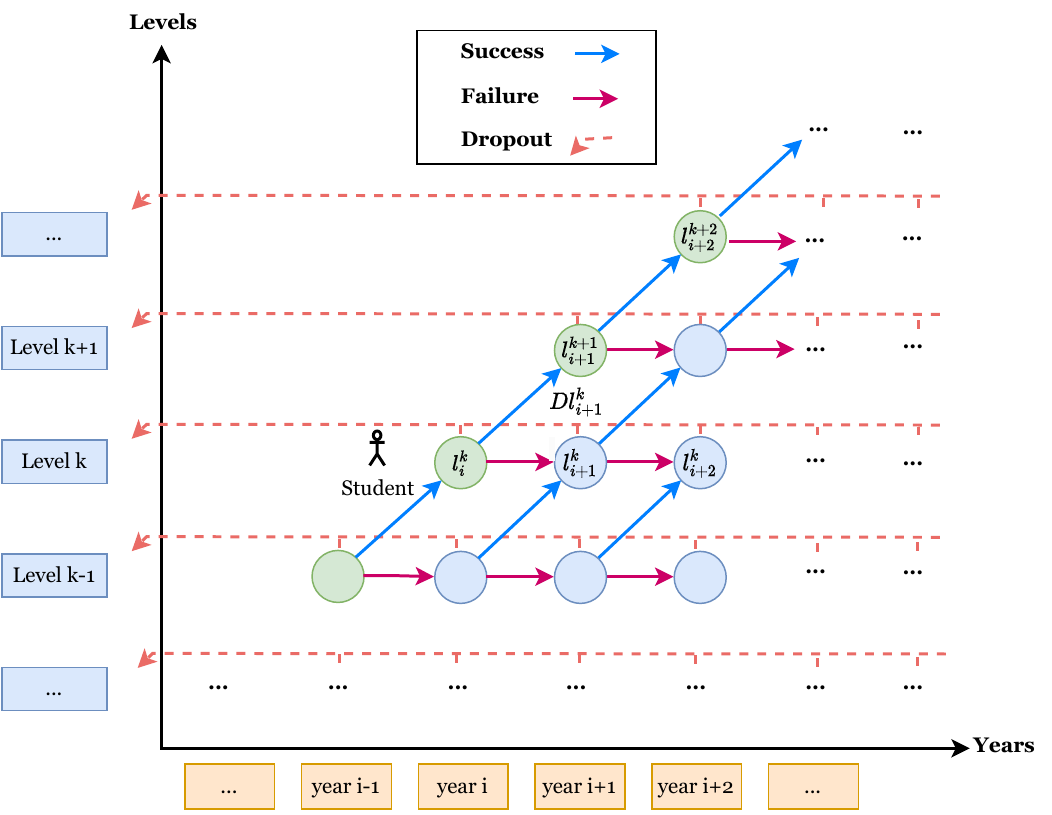}
  \caption{Student Status Tracking}
  \label{fig:Student_Status}
\end{figure}

Throughout their educational journey, students may encounter various challenges or difficulties that can lead to dropout or academic failure. Our objective is to predict the status of a student based on the data from the previous years of study for the upcoming years. The student can fall into one of three possibilities, as shown in Figure \ref{fig:Student_Status}.

\begin{itemize}
  \item \textbf{Success}: is defined as students achieving the necessary grades to progress to the subsequent level of education.
  \item \textbf{Failure}: is characterized by students not attaining the required grades to advance to the next level of education.
  \item \textbf{Dropout}: the student discontinues their education before completion.
\end{itemize}

A student enrolled at level $k$ in year $i$ is denoted by $l_{i}^k$. Subsequently, in the following year, he can dropout, denoted by $Dl_{i+1}^k$, or fail, denoted by $l_{i+1}^k$, or succeed, denoted by $l_{i+1}^{k+1}$.

\subsection{Overview of the proposed framework}

\begin{figure*}[htb!]
  \centering
  \includegraphics[width=0.9\textwidth]{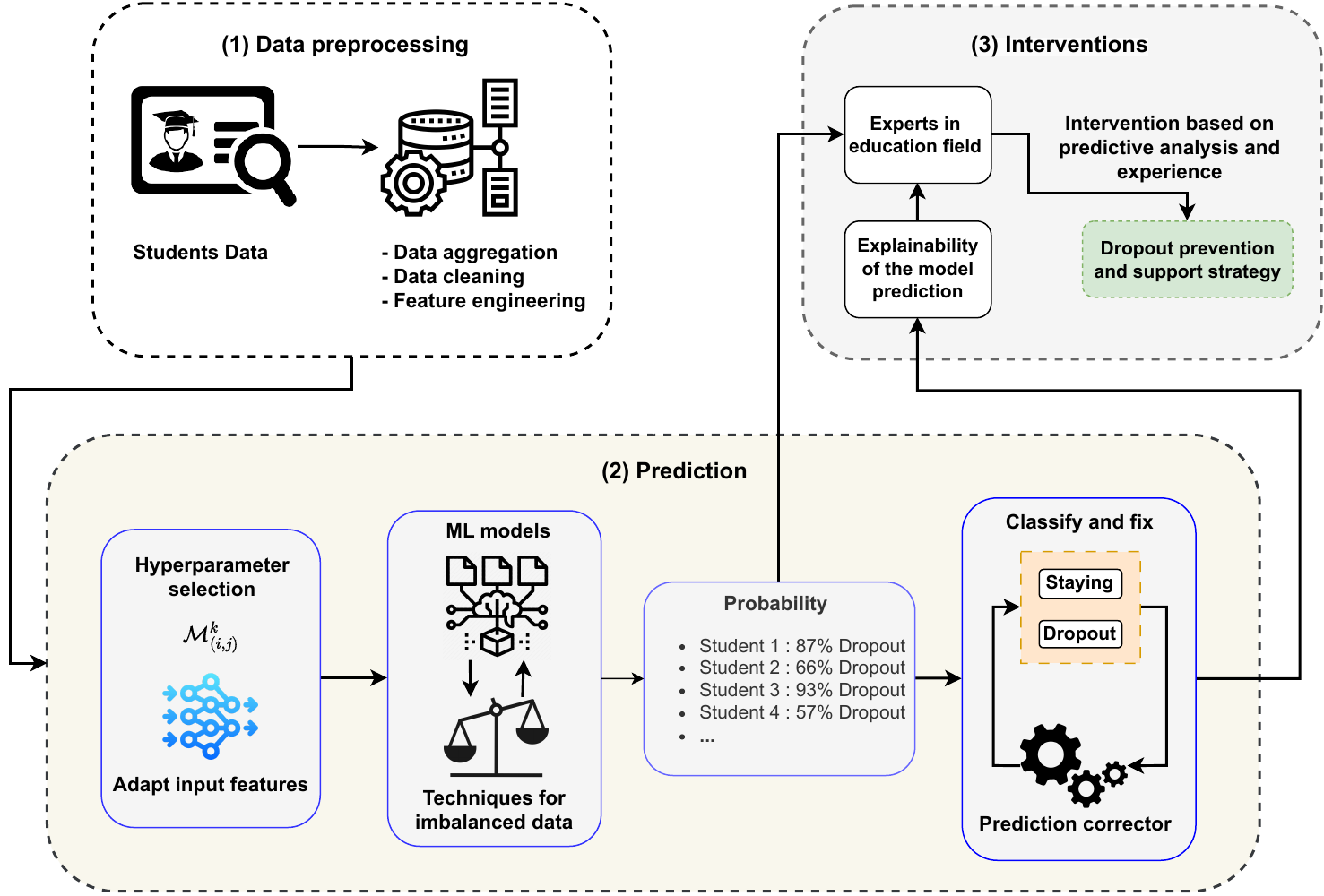}
  \caption{Overview of the proposed framework}
  \label{fig:Overview}
\end{figure*}

In this research, we present a comprehensive framework poised to make a significant contribution to the field of education. Our framework is designed to address the pressing issue of student attrition by predicting those at risk of dropout. Drawing upon an extensive historical dataset spanning one or more academic years, our approach leverages the power of machine learning. This innovative framework not only identifies students at risk in the current year but extends its predictive capabilities into the future, providing valuable insights into potential dropouts for subsequent academic years.
The proposed framework, shown in Figure \ref{fig:Overview} comprises three pivotal components: \textbf{(1) Data preprocessing}: in this preliminary stage, the data undergoes meticulous cleaning, and an array of feature engineering techniques is systematically applied, \textbf{(2) Prediction}: in this phase, we strategically select hyperparameters for our framework and curate input features for the models. We employ machine learning models and advanced techniques tailored to address the challenge of imbalanced data. Furthermore, we introduce a novel prediction corrector designed to elevate the precision of predictions within the dropout class.
\textbf{(3) Intervention}: this crucial phase is dedicated to interventions grounded in predictive analysis and practical experience, forming a cornerstone of our dropout prevention and support strategy.

\subsection{Data Preprocessing}

For the data preprocessing stage, we employ an extensive spectrum of techniques meticulously designed to optimize model performance and refine data representation \cite{L9490240I}. Our feature engineering arsenal encompasses a variety of methods, including feature extraction, which involves creating new features by grouping data based on class and school, allowing us to capture significant information. We also utilize normalization techniques, encompassing both general normalization and normalization by class. These methods ensure consistent feature scaling, preventing undue influence on our models while enhancing their convergence. Additionally, the feature encoding strategies transform categorical variables into numerical formats compatible with machine learning algorithms. This comprehensive and customized approach enables us to effectively represent categorical information. In unison, these feature engineering efforts serve as a pivotal cornerstone in our mission to enhance model accuracy and interoperability.

\subsection{Prediction}
In this section, we delve into the various stages that comprise the prediction component. Will encompass a comprehensive exploration of the stages involved in achieving accurate and effective classification, shedding light on the intricate processes that underpin this critical aspect of this study.

\subsubsection{Prediction models approach}

\begin{figure}[htb!]
  \centering
  \includegraphics[width=0.6\textwidth]{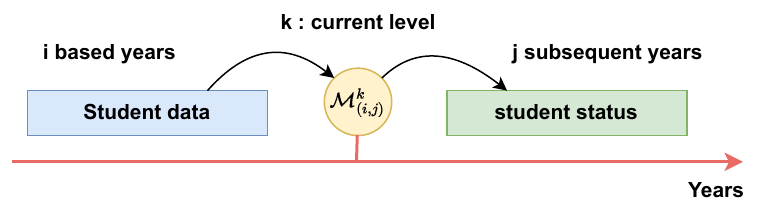}
  \caption{Predictive models approach}
  \label{fig:models_formula}
\end{figure}

Our primary aim is to develop predictive models proficient in anticipating student dropout based on historical data. As depicted in Figure \ref{fig:models_formula}, these models will be built using one or more years of historical data, enabling us to predict students at risk of dropout in upcoming academic years. This approach underscores the effectiveness and far-reaching potential of our proposed methodology, allowing us to identify at-risk students not just for the next year but for multiple years ahead.

We can represent our models with the following notation:
\[\mathcal{M}_{(i,j)}^k \]

where:
\begin{itemize}
  \item \(i\): represents the number of years used for prediction.
  \item \(j\): represents the subsequent years for which the status of students is predicted.
  \item \(k\): represents the current educational level.
\end{itemize}

The prediction models (M) used in this study include \gls{DT}, \gls{RF}, \gls{XGBoost}, \gls{LightGBM} and Ensembling, which combines these models through a voting mechanism. The choice of these models was based on their effectiveness in handling diverse datasets, providing robust predictive performance, and offering interpretability that allows for comprehensive and reliable analysis.  The proposed method is model agnostic, allowing us to explore various other machine learning models such as \gls{SVM}, \gls{LR}, \gls{KNN}, or deep learning models such as \gls{DNN}, \gls{LSTM}, \gls{CNN}, autoencoders, or even anomaly detection techniques. The choice of model depends on the characteristics of the dataset, the existing features, and the specific objectives of the study.

\subsubsection{Imbalanced Classification}

Accurate detection of dropout events is paramount in the context of machine learning (ML) models. However, achieving optimal performance can be challenging, especially when dealing with imbalanced data. To tackle this issue comprehensively, we explored various techniques to address data imbalance in our study. Specifically, we investigated the effectiveness of class weighting, undersampling, and oversampling, comparing them with the baseline approach. We employed ensemble models \cite{rodriguez2023methodology} to harness the collective strength of individual models, each of the used models was trained independently, and a voting strategy was utilized to generate the final prediction (Ensembling). This ensemble approach improved overall performance and effectively handled class imbalance in dropout detection. By comparing class weighting, undersampling, and oversampling with a baseline, we aimed to determine the most effective strategy for addressing imbalanced data and enhancing the accuracy of dropout detection in our ML models.

\subsubsection{Prediction corrector}

\begin{figure}[htb!]
  \centering
  \includegraphics[width=0.5\textwidth]{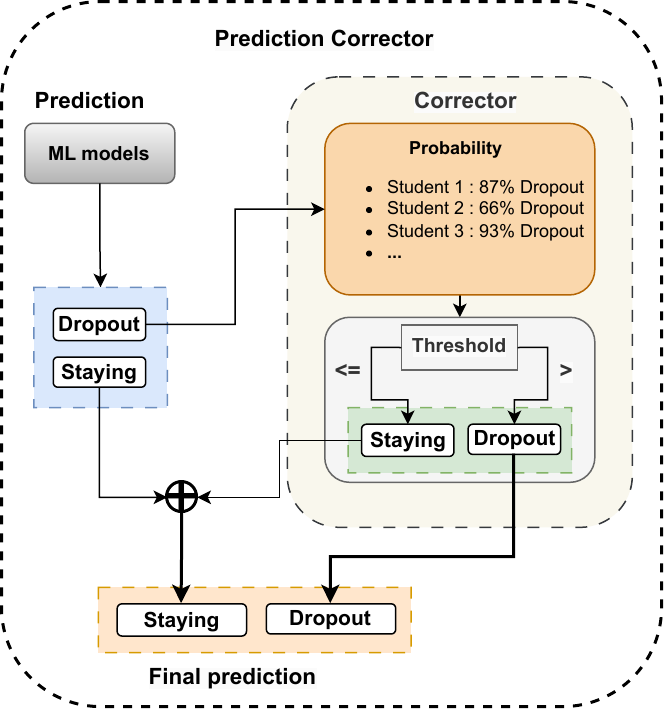}
  \caption{Prediction corrector}
  \label{fig:corrector}
\end{figure}

To improve the precision of the dropout classification, we propose a prediction corrector approach, as illustrated in Figure \ref{fig:corrector}. The principle of this corrector is as follows: when the student is predicted as a dropout, we calculate the probability of dropout for students predicted as dropouts. We then establish a threshold, and only students with a probability exceeding the threshold are considered well-predicted dropouts. We used several thresholds to assess their impact and see which one gave the most accurate prediction. 

\subsection{Interventions}
In this study, we present a novel method that harmonizes \gls{XAI} techniques with educational expertise. The XAI technique has been used in various fields, including healthcare \cite{I10322985S}, and education \cite{baranyi2020interpretable}, to enhance the transparency of models and to elucidate the specific features that influence the model's predictions. Our aim is to develop a comprehensive intervention and support strategy to proactively prevent dropout, and promote student retention. Our approach involves the use of the  \gls{SHAP} technique \cite{lundberg2017unified}, which provides insights into the most influential features affecting student dropout. This feature analysis provides a deep understanding of the intricacies surrounding dropout prediction. In addition, we aim to engage experts from the educational sector to bridge the gap between data-driven insights and practical solutions.  By uniting \gls{AI} techniques with expert knowledge, we aim to build a robust, comprehensive intervention and support strategy.

\section{Applying the proposed methodology to the Moroccan Education System}\label{sec:UseCase}

In this section, we provide an overview of the Moroccan education system and its various levels. We also detail the dataset employed in our study, its features, as well as the steps involved in data preparation. Additionally, we present key statistics related to the dataset.

\subsection{Moroccan Education System} \label{sec:Moroccan education system}

The Moroccan education system shares key elements in common with several other education systems around the world, including the United Arab Emirates \cite{gallagher2019education}, Saudi Arabia \footnote{\url{https://www.moe.gov.sa/ar/education/studies/Pages/default.aspx}}, France\footnote{\url{https://www.scholaro.com/db/countries/france/education-system}}
 , and Chile \cite{rodriguez2023methodology}. These similarities extend to fundamental aspects such as core subjects and mechanisms for student progression to subsequent levels. This underscores the global nature of educational practices and policies, emphasizing that education systems frequently draw upon international best practices and experiences.

In Morocco, the education system is a comprehensive structure that encompasses various education levels, each designed with specific objectives and curricula. This system is dedicated to furnishing students with a holistic education, equipping them with essential skills and knowledge to support their personal and professional development. The Moroccan education system's main characteristics:
    \begin{itemize}
      \item \textbf{Three-Cycle System:} Morocco's education system has a three-cycle structure, comprising primary, secondary, and higher education.
      \item \textbf{Core Subjects:} The presence of core subjects such as mathematics, science, languages, and social studies within the Moroccan education system aligns with the common curriculum components found in numerous other international education systems.
      \item \textbf{Public and Private Education:} Morocco hosts both public and private educational institutions at various levels of the education system.
      \item \textbf{National Examinations:} Morocco conducts national examinations at the end of each educational cycle, which are used to determine eligibility for the next cycle.
      \item \textbf{Different options:} Students have the opportunity to choose specialized tracks within their education, allowing them to tailor their studies to their specific interests and career goals.
    \end{itemize}

The Moroccan education system features several distinct cycles, each tailored to serve particular educational goals. Below, we outline these cycles within the Moroccan education system:

\begin{itemize}
\item \textbf{Preschool Education}: Preschool education is the first level of formal education and is not mandatory. It typically starts at the age of 5 and focuses on developing a child's social, cognitive, and motor skills through play-based activities. The goal of preschool is to prepare children for primary school and instill a love for learning.

\item \textbf{Primary Education}: Primary education is compulsory and spans six years, usually from ages six to twelve. During this phase, students receive a foundational education in various subjects, including Arabic, French, mathematics, science, social studies, and physical education. The primary education curriculum aims to develop students' literacy, numeracy, and basic knowledge.

\item \textbf{Middle School}: The middle school phase follows primary education and focuses on enhancing students' knowledge in key subjects such as mathematics, physics, and language subjects like French and English. This cycle provides students with a deeper understanding of the core subjects and lays the groundwork for more specialized studies in the future.

\item \textbf{High School}: High School marks the final stage of the educational journey. At this stage, students have the opportunity to choose from a variety of academic pathways, such as scientific, technical, and literary options, tailored to their individual interests and future career aspirations.
\end{itemize}

The transition from high school to university is a pivotal stage in students' academic journey. It is an important crossroads that gives them the freedom to define their academic path. As they enter higher education, doors open to explore fields of real passion. Certain disciplines are more selective, especially those with a high impact on society, such as medicine and engineering, where grades and knowledge are seen as the pillars on which future contributions and professional achievements will be built. The Table \ref{tab:Level} illustrates the levels of each cycle within the Moroccan education system, along with their corresponding identification codes.

    \begin{table}[htb!]
        \centering
        \caption{Levels in the Moroccan education system}
        \label{tab:Level}
        \begin{tabular}{llc}
        \hline
        Cycle & Level & ID Level \\
        \hline
        Primary & 1\textsuperscript{st} Primary Year & 1  \\
         & 2\textsuperscript{nd} Primary Year & 2  \\
         & 3\textsuperscript{th} Primary Year & 3  \\
         & 4\textsuperscript{th} Primary Year & 4 \\
         & 5\textsuperscript{th} Primary Year & 5 \\
         & 6\textsuperscript{th} Primary Year & 6 \\
        \midrule
        Middle School & 1\textsuperscript{st} Middle School Year & 7  \\
         & 2\textsuperscript{nd} Middle School Year & 8 \\
         & 3\textsuperscript{rd} Middle School Year & 9 \\
        \midrule
        High School & Common Core & 10 \\
         & 1\textsuperscript{st} Year Bac & 11 \\
         & 2\textsuperscript{nd} Year Bac & 12 \\
        \hline
        \end{tabular}
    \end{table}

In 2021, Morocco's exceptional commitment to educational advancement was acknowledged on the global stage when it secured 57\textsuperscript{th} position in worldwide education rankings\footnote{\url{https://worldpopulationreview.com/country-rankings/education-rankings-by-country}}
.  This achievement reflects Morocco's dedicated efforts to enhance its education system and underscores its commitment to providing accessible and quality education to its citizens. With a rich cultural heritage and a growing emphasis on education, Morocco continues to make strides towards improving educational outcomes and opportunities for its students. This ranking serves as a testament to the nation's commitment to nurturing the potential of its youth and ensuring a brighter future for generations to come.

\subsection{Dataset}\label{sec:Dataset}
\begin{figure}[htb!]
  \centering
  \includegraphics[width=0.6\textwidth]{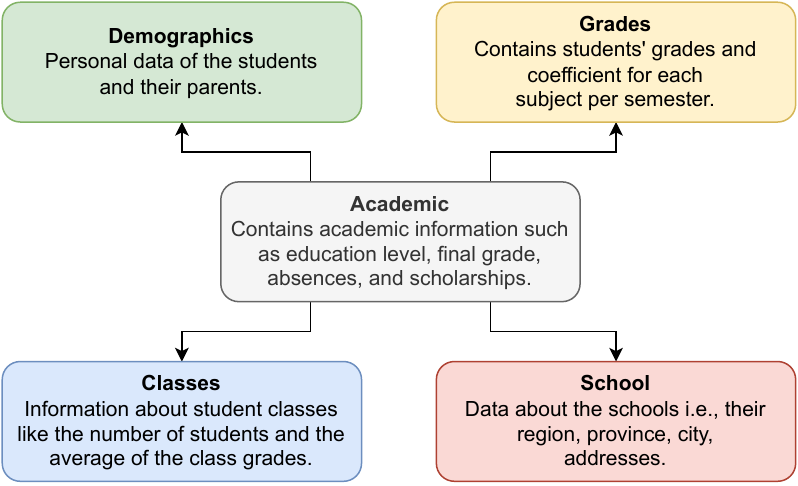}
  \caption{Dataset architecture}
  \label{fig:dataset}
\end{figure}

The dataset utilized in this research, as illustrated in Figure \ref{fig:dataset} is stored in a relational database provided by the Moroccan Ministry of National Education, Preschool, and Sports. It includes a wide range of information, including academic records, student demographic/personal data, and additional data related to teachers, schools, and classes. This comprehensive dataset allows us to effectively explore and analyze the multitude of factors that may influence dropout rates. The main aim is to gain valuable insights into the causes of dropout and to propose effective intervention strategies.

\begin{table*}[htb!]
  \small 
  \centering
  \caption{Features of the dataset}
  \label{tab:Features}
  \resizebox{1\hsize}{!}{$
    \begin{tabular}{|c|c|c|c|}
    \hline
    \textbf{Category} & \textbf{Features} & \textbf{Range of features}  & \textbf{Missing values} \\
    \hline
    \textbf{Academic} & Year of education & 2015/2016 to 2020/2021 & 0\% \\
    & Participation in the Cartable program (It is a scholarship) & Binary 0 or 1 & 0\%  \\
    & Participation in the Tayssir program (It is a scholarship) & Binary 0 or 1 & 0\% \\
    & Overall grades average & From 0 to 20 & 2.27\% \\
    & Number of days missed (authorized) & From 0 to 120 & 1.68\% \\
    & Number of classes missed (authorized) & From 0 to 828 & 30.38\% \\
    & Number of days missed (unauthorized) & From 0 to 388 & 30.38\% \\
    & Number of classes missed (unauthorized)  & From 0 to 101 & 30.38\% \\
    & Number of years with failures at the current level & From 0 to 3 & 0\% \\
    & Ranking in class  & From 0 to 50 & 0\% \\
    \hline
    \textbf{Grade} & Average grades for each subject &  From 0 to 20& 1.59 \% \\
    & Coefficient of each subject &  From 1 to 6 & 1.59 \% \\  
    & Average of  scientific subjects &  From 0 to 20& 0\% \\
    & Average of  literary subjects &  From 0 to 20& 0\% \\
    \hline
    \textbf{Demographic} & Student gender & Binary 0 or 1  & 1.59\% \\
    & Student nationality & Binary 0 or 1 & 17.84\% \\
    & Student birthplace & 121720 distinct values & 36.92\% \\
    & Presence of a disability & 6 distinct values & 0\% \\
    & Attendance at preschool  & 3 distinct values & 35.81\% \\
    & Father's profession  & 16849 distinct values & 47.08\% \\
    & Mother's profession  & 5462 distinct values & 63.58\% \\
    & Age at current academic level & From 6 to 23 & 0\% \\
    \hline
    \textbf{Class} &  Number of students in the class & From 1 to 121& 0\% \\
    & Number of female students in the class & From 0 to 61& 0\% \\
    & Mean grade of the class & From 0 to 20 & 0\% \\
    & Number of student failure in the class & From 0 to 76  & 0\% \\
    & Number of students dropped in the class & From 0 to 52 & 0\% \\
    & Number of levels in the class & From 0 to 6 & 0\% \\
    
    \hline
    \textbf{Schools} & Number of years since the opening of the establishment & From 2 to 88 & 39.48 \% \\
    & Province  & 9 distinct values & 0\% \\
    & Boarding school status & Binary 0 or 1 & 0\% \\
    & Availability of internet & Binary 0 or 1 & 0\% \\
    & School city & 407 distinct values & 26.87\% \\
    & Tayssir program participation & Binary 0 or 1 & 0\% \\
    & Number of student failure in the school & From 0 to 1021 & 0\% \\
    & Number of students dropped in the school & From 0 to 850 & 0\% \\
    
    \hline
  \end{tabular}$}
\end{table*}

\begin{figure}[htb!]
  \centering
  \includegraphics[width=0.6\textwidth]{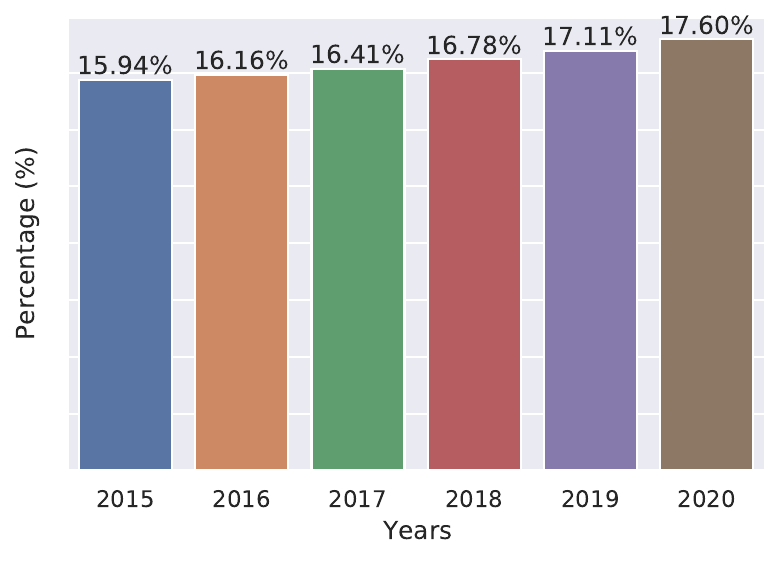}
  \caption{Distribution of the number of students by academic year in the dataset: total summing to 100\%}
  \label{fig:dataset_distribution}
\end{figure}

Morocco consists of 12 regions,  each containing many cities and villages, this study draws on a comprehensive dataset that provides a detailed insight into the educational landscape of the Fez-Meknes region. This unique dataset encompasses an extensive collection of records, spanning five consecutive academic years, and a remarkable 1.4 million student profiles. With a rich array of more than 100 features, in this study, we selected 37 distinct features as shown in the Table \ref{tab:Features}. Our dataset offers a multifaceted perspective on various aspects of the educational journey. In Figure \ref{fig:dataset_distribution}, we present an overview of the distribution of the dataset across academic years, i.e., the number of students in each academic year.

\subsubsection{Data Preparation}
In this step, our primary objective is to improve the quality of the data through various actions such as joining tables and addressing different issues present in the dataset, particularly missing values.
First, we focus on the process of merging tables. This involves combining several tables that contain related information about the same student, using the primary keys of the tables. This allows us to gain a more holistic view and conduct a more insightful analysis. Additionally, we need to tackle the problem of missing values within the dataset. To overcome this challenge, we use techniques such as imputation, which involves filling in missing values using a combination of statistical methods and expert judgment. Furthermore, during the data preparation process, we may encounter other issues that require attention. These issues include inconsistent data formats, duplicate records, or errors in the dataset. By addressing all of these issues, we aim to ensure the quality of the data and make it suitable for subsequent tasks such as analysis, modeling, or any other desired purpose.

\subsubsection{Data Labeling}
Data labeling is an essential step in supervised machine learning problems. In our specific case, we have two tables that provide labels for each student. The first table contains information about the student's current situation, indicating whether they have dropped out or are continuing their studies. The second table contains the students' final results, indicating whether they were successful or not. By combining these tables, we can determine whether a student is a dropout or not based on the first table, and if they are still studying, whether they are a success or not based on the second table. This merging allows us to create a comprehensive dataset that covers both the student's current situation and their result.

\begin{table*}[htb!]
\centering
\caption{Evolution of the dropout rate by cycle of education (\%)}
\label{tab:cycle_dropout_rate}

\begin{tabular}{lcccccc}
\hline
Cycle & 2015/2016 & 2016/2017 & 2017/2018 & 2018/2019 & 2019/2020 & 2020/2021 \\
\hline
Primary & 2.54 & 2.56 & 2.46 & 1.96 & 1.77 & 2.45 \\
\textbf{Middle School} & \textbf{13.95} & \textbf{14.74} & \textbf{14.41} & \textbf{13.58} & \textbf{9.20} & \textbf{13.62} \\
High School & 11.20 & 11.30 & 12.65 & 12.35 & 7.13 & 7.67 \\
Overall & 6.80 & 7.02 & 7.04 & 6.52 & 4.48 & 6.12 \\
\hline
\end{tabular} 
\end{table*}

\begin{table*}[htb!]
\centering
\caption{Evolution of the dropout rate by level of education (\%)}
\label{tab:level_dropout_rate}
\resizebox{1\hsize}{!}{$
\begin{tabular}{llcccccc}
\hline
Cycle & Level & 2015/2016 & 2016/2017 & 2017/2018 & 2018/2019 & 2019/2020 & 2020/2021 \\
\hline
Primary & 1\textsuperscript{st} Primary Year & 2.42 & 2.93 & 2.91 & 3.08 & \textbf{3.73} & \textbf{5.87} \\
 & 2\textsuperscript{nd} Primary Year & 1.07 & 1.07 & 1.09 & 1.12 & 1.01 & 1.45 \\
 & 3\textsuperscript{rd} Primary Year & 0.95 & 1.11 & 1.13 & 1.08 & 1.03 & 1.95 \\
 & 4\textsuperscript{th} Primary Year & 1.62 & 1.43 & 1.41 & 1.13 & 0.83 & 1.01 \\
 & 5\textsuperscript{th} Primary Year & 2.41 & 2.45 & 2.4 & 1.75 & 1.28 & 1.55 \\
 & \textbf{6\textsuperscript{th} Primary Year} & \textbf{6.49} & \textbf{6.28} & \textbf{5.87} & \textbf{3.68} & 2.49 & 2.24 \\
\midrule
Middle School & 1\textsuperscript{st} Middle School Year & 11.74 & 13.52 & 14.1 & 13.88 & \textbf{10.98} & \textbf{15.32} \\
 & 2\textsuperscript{nd}Middle School Year & 9.84 & 11.46 & 11.43 & 10.51 & 7.82 & 11.11 \\
 & \textbf{3\textsuperscript{rd} Middle School Year} & \textbf{19.41} & \textbf{18.96} & \textbf{17.46} & \textbf{16.1} & 8.35 & 14.01 \\
\midrule
High School & Common Core & 6.99 & 8.23 & 7.94 & 7.41 & 5.83 & 6.37 \\
 & 1\textsuperscript{st} Year Bac & 8.99 & 8.47 & 8.43 & 7.67 & 5.6 & 6.28 \\
 & \textbf{2\textsuperscript{nd} Year Bac} & \textbf{17.69} & \textbf{17.49} & \textbf{21.22} & \textbf{20.32} & \textbf{10.14} & \textbf{11.1} \\
\hline
\end{tabular}
$}
\end{table*}

\subsubsection{Dropout analysis}
To gain comprehensive insights into the factors contributing to student dropout, we undertook a thorough analysis. Our primary objective is to calculate the dropout rate by taking into account several factors, including the educational cycle, the level of study, and the academic years. This comprehensive approach allows us to paint a clearer picture of the dropout phenomenon and provides valuable information for developing targeted interventions and strategies to mitigate dropout rates effectively. The result obtained is displayed in two tables, Table \ref{tab:cycle_dropout_rate} shows the evolution of the dropout rate by the cycle of education, while Table \ref{tab:level_dropout_rate}  presents the evolution of the dropout rate by level of education.

The analysis of dropout rates reveals remarkable results. In particular, as shown in Table \ref{tab:cycle_dropout_rate}, the middle school cycle emerges as the cycle with the highest dropout rate, closely followed by the high school cycle. These observations suggest that students passing through middle or high school may face challenges that contribute to dropping out. In contrast,  the primary cycle has the lowest dropout rate, suggesting a comparatively more stable situation for students in this cycle. In addition, as shown in Table \ref{tab:level_dropout_rate}, the data show that the peaks in dropout rates occur in the final years of each cycle.  Such pivotal points require careful attention to provide students with appropriate support and resources to enable them to complete their education successfully.

\section{Experiments and Results}\label{sec:Experiments}

\subsection{Metrics}

To thoroughly evaluate models and gain a comprehensive understanding of their performance, we have employed various metrics. By assessing multiple evaluation metrics, including accuracy, recall, precision, and F1-score, we aim to obtain a holistic view of the model's effectiveness.

For classification, prediction outcomes belong to one of the four cases:
\begin{itemize}
  \item \gls{TP} corresponds to positive samples that are correctly classified as positive.
  \item \gls{TN} indicates samples that are actually negative and are correctly classified as negative.
  \item \gls{FP} refers to negative samples incorrectly classified.
  \item \gls{FN} refers to positive samples incorrectly classified.
\end{itemize}

Based on these outcomes, evaluation metrics are defined as given below:

\begin{itemize}
    \item \textbf{Accuracy}: is a measure of correct predictions over total predictions:
    $$Accuracy=\frac{TP+TN}{TP+TN+FP+FN}$$

    \item \textbf{Precision}: measures the proportion of TP among the samples classified as positive: 
    $$Precision=\frac{TP}{TP+FP}$$
    \item \textbf{Recall}: measures the proportion of actual positives that are correctly identified:
    $$Recall=\frac{TP}{TP+FN}$$
    \item \textbf{F1-score}: is a weighted average of Precision and Recall: 
    $$F1 =2 \times \frac{Recall \times Precision}{Recall+Precision}$$
    \item \textbf{AUC}: \acrfull{AUC} evaluates the effectiveness of binary classification models by quantifying the area under the \gls{ROC} curve, a visual indicator of model performance. The ROC curve illustrates the model's ability to distinguish true positives  (sensitivity or recall) from false positives (1 - specificity) as the threshold is varied.
    $$Specificity=\frac{TN}{TN+FP}$$
\end{itemize}

By considering these metrics collectively, we can comprehensively evaluate the model's performance and make informed decisions about its effectiveness for the given task.

\subsection{Splitting the Dataset}
To ensure a thorough evaluation of our models, we propose to assess their performance using multiple test plans. These test plans allow us to gain a comprehensive understanding of how well the models generalize and how effectively they identify students at risk of dropping out.

\begin{itemize}
  \item \textbf{Guided random split:} According to our research, it is commonly observed that many articles use a random split to evaluate the models \cite{song2023all}. However, to ensure that the test dataset contains student data from all academic years, we propose a different approach. Specifically, as shown in Figure \ref{fig:Split_random}, From each academic year, we randomly select 20\% of the data for testing. This approach helps to ensure that the test dataset is representative of the full range of academic years and takes into account any potential variations or patterns specific to different academic years.

  \item \textbf{Split by schools:} We propose to split the data by school, specifically as shown in Figure \ref{fig:Split_byschools}, by selecting schools from different areas. This approach is intended to ensure that the models are trained and evaluated in a variety of educational contexts, taking into account potential variations between different regions and schools. By splitting the data by school, we can capture many factors that may influence student outcomes, such as differences in resources, teaching methods, or socioeconomic factors.

  


  \begin{figure}[htb!]
    \centering
    \begin{subfigure}[b]{0.4\textwidth}
        \centering
        \includegraphics[width=\textwidth]{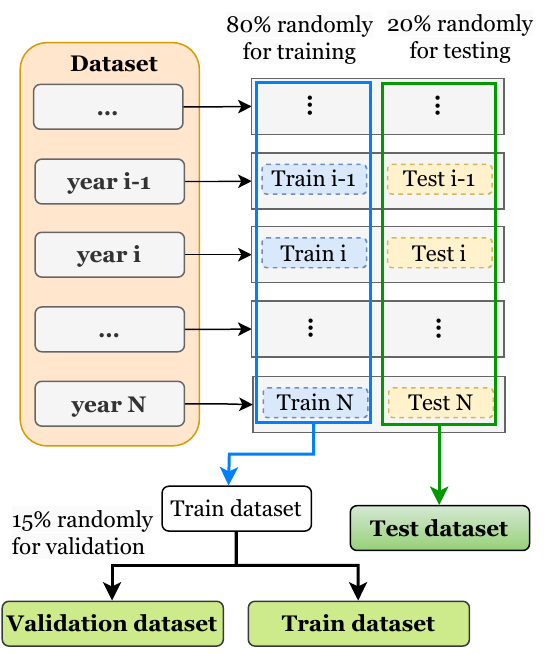}
        \caption{Guided random split}
        \label{fig:Split_random}
    \end{subfigure}
    \quad 
    \begin{subfigure}[b]{0.5\textwidth}
        \centering
        \includegraphics[width=\textwidth]{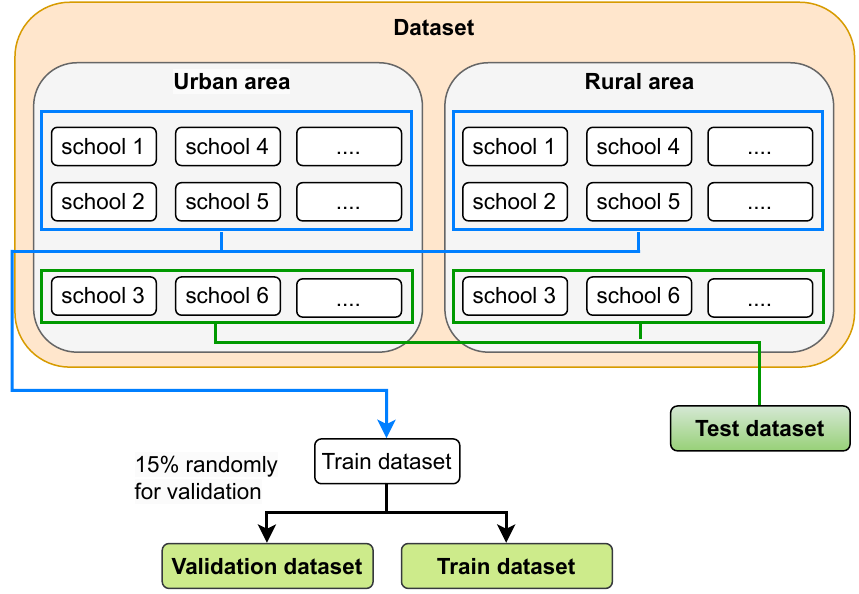}
        \caption{Split by schools}
        \label{fig:Split_byschools}
    \end{subfigure}
    \quad
    \begin{subfigure}[b]{0.6\textwidth}
        \centering
        \includegraphics[width=\textwidth]{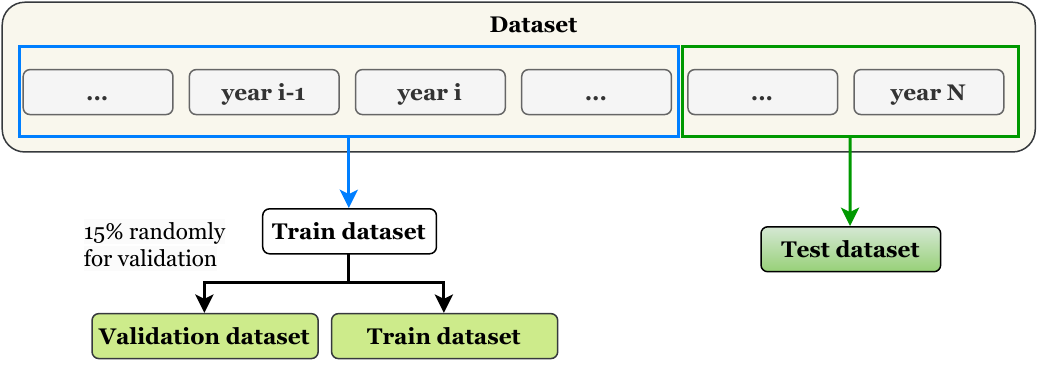}
        \caption{Split by years}
        \label{fig:Split_byyears}
    \end{subfigure}
    
  \caption{Different split approaches}
    \label{fig:Comparison}
\end{figure}

  \item  \textbf{Split by years:} We propose to split the data based on academic year as shown in Figure \ref{fig:Split_byyears}, where we train our model on the previous years' data and reserve the most recent (j) years for testing. This approach provides a realistic evaluation of the model's performance as it simulates the deployment scenario where the model has to predict outcomes for students from the new academic year (j).
\end{itemize}

\subsection{Comparative analysis of imbalanced data techniques}
To address the challenges posed by imbalanced datasets, a comparative analysis of various techniques has been conducted. This analysis aims to delve into and juxtapose the performance of distinct strategies in managing data imbalances. For this purpose, four specific techniques have been meticulously selected for evaluation: the baseline approach, class weighting, oversampling, and undersampling. The Table \ref{tab:comparison} presents a succinct depiction of the outcomes derived from the comparison among these techniques. Here are some notes on the results of the comparisons:

\begin{table*}[htb!]
  \centering
  \caption{Comparison of imbalanced data techniques for \( \mathcal{M}_{(1,1)}^6 \)}
  \label{tab:comparison}
  \resizebox{1\hsize}{!}{$
  \begin{tabular}{|c|c|c|c|c|c|c|c|c|}
  \hline
    Technique & Algorithm & Accuracy & \multicolumn{2}{c|} {Class 0 (Continue)} & \multicolumn{2}{c|}{Class 1 (Dropout)} & F1-Score & AUC \\ 

     &&& Recall & Precision & Recall & Precision & & \\
    \hline

    Baseline & Decision Tree & 0.92 & 0.99 & 0.93 & 0.18  & 0.60 &  0.62 & 0.58 \\

    & Random Forest & 0.92 & 0.99 & 0.93 & 0.28 & 0.66 & 0.68 & 0.62 \\
    & XGBoost & 0.92 & 0.99 & 0.93 & 0.28 & 0.66 & 0.68 & 0.63 \\
    & \textbf{LightGBM} & \textbf{0.93} & \textbf{0.99} & \textbf{0.93} & \textbf{0.28 }& \textbf{0.66} & \textbf{0.68} &\textbf{0.63} \\
    & Ensembling & 0.92 & 0.99 & 0.93 & 0.25 & 0.69 & 0.66 & 0.63 \\
    \hline
    Class weights & Decision Tree & 0.72 & 0.71 & 0.98 & 0.84 & 0.22 & 0.59 & 0.77 \\
    & Random Forest & 0.73 & 0.72 & 0.98 & 0.84 & 0.22 & 0.59 & 0.77 \\
    & XGBoost & 0.80 & 0.89 & 0.98 & 0.84 & 0.28 & 0.65 & 0.81 \\
    & \textbf{LightGBM} & \textbf{0.80} & \textbf{0.80} & \textbf{0.98} & \textbf{0.84} & \textbf{0.29} & \textbf{0.65} & \textbf{0.81} \\
    & Ensembling & 0.79 & 0.78 & 0.98 & 0.84 & 0.28 & 0.65 & 0.81 \\
    \hline
    \textbf{Undersampling} & Decision Tree & 0.76 & 0.76 & 0.98 & 0.80 & 0.24 & 0.61 & 0.78 \\
    & Random Forest & 0.77 & 0.77 & 0.98 & 0.82 & 0.26 & 0.63 & 0.79 \\
    & XGBoost & 0.80 & 0.80 & 0.98 & 0.83 & 0.29 & 0.65 & 0.81 \\
    & \textbf{LightGBM} & \textbf{0.81} & \textbf{0.80} & \textbf{0.98} & \textbf{0.83} & \textbf{0.29} & \textbf{0.65} & \textbf{0.81} \\
    & Ensembling & 0.80 & 0.80 & 0.98 & 0.84 & 0.28 & 0.65 & 0.81 \\
    \hline
    Oversampling & Decision Tree & 0.85 & 0.90 & 0.94 & 0.38 & 0.27 & 0.62 & 0.63 \\
    (SMOTE) & \textbf{Random Forest} & \textbf{0.85} & \textbf{0.88 }& \textbf{0.96} & \textbf{0.63} & \textbf{0.33} & \textbf{0.67 }& \textbf{0.75} \\
    & XGBoost & 0.92 & 0.98 & 0.94 & 0.31 & 0.63 & 0.69 & 0.64 \\
    & LightGBM & 0.92 & 0.98 & 0.94 & 0.30 & 0.64 & 0.68 & 0.64 \\
    & Ensembling & 0.92 & 0.98 & 0.94 & 0.33 & 0.62 & 0.70 & 0.64 \\
    \hline
  \end{tabular} $}
\end{table*}

\begin{itemize}
  \item \textbf{Baseline}: The baseline models \gls{DT}, \gls{RF}, \gls{XGBoost}, \gls{LightGBM}, and Ensembling show relatively high accuracy and recall for class 0 (Continue) but struggle with class 1 (Dropout) prediction, resulting in lower recall, and precision for the dropout class.
  \item \textbf{Class weights}: Class weighting leads to better performance for class 1 (Dropout) prediction, as indicated by the improved recall of the dropout class. However, this is at the expense of reduced precision of the dropout class.
  \item \textbf{Undersampling}: Reducing the number of samples in the majority class (Continue) to equal the number of samples in the dropout class helps to improve the performance of dropout class prediction by increasing recall. However, it may result in a slight decrease in the accuracy of the models and also in the recall of the dropout class.
  \item \textbf{Oversampling (SMOTE)}: The oversampling of the minority class (Dropout) using the SMOTE technique \cite{chawla2002smote}, aimed to increase the number of instances of class 1 to improve the performance of the models for this class, but the results show that it did not provide the expected improvements compared to the other techniques such as class weights and undersampling.
\end{itemize}

The \textbf{LightGBM} algorithm, especially when used with class weights, and the undersampling techniques, appears to be the best-performing model overall. It achieves the highest recall, precision, F1-score, and AUC for the dropout class while maintaining reasonable performance for class 0.

\subsection{Results of  $\mathcal{M}_{(1,1)}^k$ Models}

\begin{table*}[htb!]
  \centering
  \caption{Performance of the models $\mathcal{M}_{(1,1)}^k$ using the guided random split approach}
  \label{tab:results_random}
  \resizebox{1\hsize}{!}{$
  \begin{tabular}{|c|c|c|c|c|c|c|c|c|}
    \hline
    Level & Algorithm & Accuracy & \multicolumn{2}{c|} {Class 0 (Continue)} & \multicolumn{2}{c|}{Class 1 (Dropout)} & F1-Score & AUC \\ 

     & & & Recall & Precision & Recall & Precision & & \\
    \hline
      \( \mathcal{M}_{(1,1)}^5 \) & Decision Tree & 0.90 & 0.91 & 0.99 & 0.62 & 0.15 & 0.60 & 0.76 \\
    & Random Forest & 0.91 & 0.91 & 0.99 & 0.64 & 0.16 & 0.60 & 0.77 \\
    & XGBoost & 0.90 & 0.90 & 0.99 & 0.71 & 0.16 & 0.60 & 0.80 \\
    & \textbf{LightGBM} & \textbf{0.90} & \textbf{0.90} & \textbf{0.99} & \textbf{0.73} & \textbf{0.16} & \textbf{0.60} & \textbf{0.81} \\
    & Ensembling & 0.91 & 0.91 & 0.99 & 0.70 & 0.16 & 0.61 & 0.81 \\
    \hline
    \( \mathcal{M}_{(1,1)}^6 \) & Decision Tree & 0.76 & 0.76 & 0.98 & 0.81 & 0.24 & 0.61 & 0.78 \\
    & Random Forest & 0.77 & 0.77 & 0.98 & 0.82 & 0.26 & 0.63 & 0.79 \\
    & XGBoost & 0.81 & 0.80 & 0.98 & 0.84 & 0.29 & 0.66 & 0.82 \\
    & \textbf{LightGBM} & \textbf{0.81} & \textbf{0.80} & \textbf{0.98} & \textbf{0.84} & \textbf{0.29} & \textbf{0.66} & \textbf{0.82} \\
    & Ensembling & 0.80 & 0.80 & 0.98 & 0.84 & 0.29 & 0.66 & 0.81 \\
    \hline
    \( \mathcal{M}_{(1,1)}^7 \) & Decision Tree & 0.82 & 0.81 & 0.97 & 0.86& 0.45 &  0.74 & 0.83 \\
    & Random forest & 0.80 & 0.79 & 0.98 & 0.89& 0.43 &  0.73 & 0.84 \\
    & XGBoost & 0.83 & 0.82 & 0.98 & 0.89& 0.45 &  0.75 & 0.85 \\
    & \textbf{LightGBM}     & \textbf{0.83} & \textbf{0.82} & \textbf{0.98} & \textbf{0.89}& \textbf{0.46} &  \textbf{0.75} & \textbf{0.85} \\
    & Ensembling    & 0.82 & 0.81 & 0.98 & 0.89& 0.46 &  0.75 & 0.85 \\
    \hline
    \( \mathcal{M}_{(1,1)}^8 \) & Decision Tree & 0.78 & 0.77 & 0.97 & 0.87& 0.36 &  0.69 & 0.81 \\
    & Random forest & 0.78 & 0.76 & 0.98 & 0.88& 0.36 &  0.68 & 0.81 \\
    & XGBoost & 0.81 & 0.80 & 0.98 & 0.87& 0.39 &  0.71& 0.83 \\
    & \textbf{LightGBM}      & \textbf{0.81} &\textbf{0.82} & \textbf{0.98} & \textbf{0.87}&\textbf{0.39} &  \textbf{0.71} & \textbf{0.83} \\
    & Ensembling    & 0.80 & 0.79 & 0.98 & 0.88& 0.39 &  0.70 & 0.83 \\
    \hline
    \( \mathcal{M}_{(1,1)}^9 \) & Decision Tree & 0.83 & 0.84 & 0.96 & 0.77& 0.42 &  0.72 & 0.80 \\
    & Random forest & 0.82 & 0.82 & 0.96 & 0.79& 0.40 &  0.71 & 0.80 \\
    & XGBoost       & 0.83 & 0.83 & 0.97 & 0.81& 0.42 &  0.72 & 0.82 \\
    & \textbf{LightGBM}      & \textbf{0.83} & \textbf{0.83} & \textbf{0.97} & \textbf{0.81}& \textbf{0.42} &  \textbf{0.72} & \textbf{0.82} \\
    & Ensembling    & 0.83 & 0.83 & 0.97 & 0.81& 0.42 &  0.73 & 0.82 \\
    \hline
  \end{tabular} 
  $}
\end{table*}

\begin{table*}[htb!]
  \centering
  \caption{Performance of the models $\mathcal{M}_{(1,1)}^k$ using the split by schools approach}
  \label{tab:results_school}
  \resizebox{1\hsize}{!}{$
  \begin{tabular}{|c|c|c|c|c|c|c|c|c|}
    \hline
    Level & Algorithm & Accuracy & \multicolumn{2}{c|} {Class 0 (Continue)} & \multicolumn{2}{c|}{Class 1 (Dropout)} & F1-Score & AUC \\ 

     &&& Recall & Precision & Recall & Precision & & \\
    \hline
    \( \mathcal{M}_{(1,1)}^5 \) & Decision Tree & 0.93 & 0.94 & 0.99 & 0.64& 0.19 &  0.63 & 0.78 \\
    & Random forest & 0.93 & 0.93 & 0.99 & 0.65& 0.18 &  0.62 & 0.79 \\
    & XGBoost & 0.93 & 0.93 & 0.99 & 0.65& 0.18 & 0.63 & 0.80 \\
    & \textbf{LightGBM} & \textbf{0.93} & \textbf{0.93} & \textbf{0.99} & \textbf{0.69} & \textbf{0.18} &  \textbf{0.63} & \textbf{0.80} \\
    & Ensembling & 0.93 & 0.94 & 0.99 & 0.68 & 0.19 &  0.63 & 0.80 \\
    \hline
    \( \mathcal{M}_{(1,1)}^6 \) & Decision Tree & 0.89 & 0.90 & 0.98 & 0.66& 0.23 &  0.64 & 0.77 \\
    & Random forest & 0.91 & 0.93 & 0.98 & 0.56& 0.27 &  0.66 & 0.74 \\
    & \textbf{XGBoost}       & \textbf{0.90} & \textbf{0.91} & \textbf{0.98} & \textbf{0.70}& \textbf{0.25} & \textbf{0.66} & \textbf{0.80} \\
    & LightGBM      & 0.90 & 0.91 & 0.98 & 0.69& 0.25 &  0.66 & 0.79 \\
    & Ensembling    & 0.90 & 0.91 & 0.98 & 0.68& 0.26 &  0.66 & 0.79 \\
    \hline
    \( \mathcal{M}_{(1,1)}^7 \) & Decision Tree & 0.81 & 0.79 & 0.98 & 0.88& 0.41 &  0.72 & 0.83 \\
    & Random forest & 0.81 & 0.80 & 0.98 & 0.88& 0.42 &  0.72 & 0.83\\
    & XGBoost       & 0.83 & 0.83 & 0.97 & 0.87& 0.45 &  0.74 & 0.84 \\
    & \textbf{LightGBM}      & \textbf{0.83} & \textbf{0.83} & \textbf{0.97} & \textbf{0.87}& \textbf{0.45} &  \textbf{0.74} & \textbf{0.84} \\
    & Ensembling    & 0.83 & 0.82 & 0.98 & 0.88& 0.45 &  0.74 & 0.84 \\
    \hline
   \( \mathcal{M}_{(1,1)}^8 \) & Decision Tree & 0.80 & 0.79 & 0.97 & 0.84& 0.40 &  0.71 & 0.81 \\
    & Random forest & 0.78 & 0.77 & 0.97 & 0.88& 0.38 &  0.70 & 0.82 \\
    & XGBoost       & 0.80 & 0.80 & 0.97 & 0.86& 0.41 &  0.72 & 0.82 \\
    & LightGBM      & 0.81 & 0.80 & 0.97 & 0.85& 0.41 &  0.72 & 0.82 \\
    & \textbf{Ensembling}    & \textbf{0.80} & \textbf{0.79} & \textbf{0.97} & \textbf{0.87}& \textbf{0.41} &  \textbf{0.71} & \textbf{0.82} \\
    \hline
    \( \mathcal{M}_{(1,1)}^9 \) & Decision Tree & 0.81 & 0.80 & 0.96 & 0.82& 0.40 &  0.71 & 0.81 \\
    & Random forest & 0.81 & 0.81 & 0.96 & 0.81& 0.42 &  0.72 & 0.81 \\
    & XGBoost       & 0.83 & 0.83 & 0.97 & 0.82& 0.44 &  0.73 & 0.82 \\
    & LightGBM      & 0.83 & 0.83 & 0.96 & 0.81& 0.44 &  0.73 & 0.82 \\
    & \textbf{Ensembling}    & \textbf{0.83} & \textbf{0.83} & \textbf{0.97}& \textbf{0.82}& \textbf{0.45} &  \textbf{0.74} & \textbf{0.82} \\
    \hline
  \end{tabular}$}
\end{table*}

\begin{table*}[htb!]
  \centering
  \caption{Performance of the models $\mathcal{M}_{(1,1)}^k$ using the split by years approach}
  \label{tab:results_years}
  \resizebox{1\hsize}{!}{$
  \begin{tabular}{|c|c|c|c|c|c|c|c|c|}
    \hline
    Level & Algorithm & Accuracy & \multicolumn{2}{c|} {Class 0 (Continue)} & \multicolumn{2}{c|}{Class 1 (Dropout)} & F1-Score & AUC \\ 
     &&& Recall & Precision & Recall & Precision & & \\
    \hline
    \( \mathcal{M}_{(1,1)}^5 \) & Decision Tree & 0.89 & 0.90 & 0.99 & 0.58& 0.12 &  0.57 & 0.73 \\
    & Random forest & 0.88 & 0.89 & 0.99 & 0.63 & 0.12 &  0.57 & 0.75 \\
    & XGBoost       & 0.91 & 0.92 & 0.99 & 0.61 & 0.15 &  0.60 & 0.76 \\
    & LightGBM      & 0.89 & 0.89 & 0.99 & 0.65 & 0.13 &  0.58 & 0.77 \\
    & \textbf{Ensembling}    & \textbf{0.90} & \textbf{0.91} & \textbf{0.99} & \textbf{0.63} & \textbf{0.14} &  \textbf{0.60} & \textbf{0.77} \\
    \hline
    \( \mathcal{M}_{(1,1)}^6 \) & Decision Tree & 0.75 & 0.75 & 0.97 & 0.73& 0.22 &  0.60 & 0.74 \\
    & Random forest & 0.78 & 0.87 & 0.97 & 0.70& 0.24 &  0.61 & 0.74 \\
    & XGBoost       & 0.85 & 0.78 & 0.96 & 0.63& 0.31 &  0.66 & 0.74 \\
    & LightGBM      & 0.84 & 0.86 & 0.96 & 0.63& 0.30 &  0.66 & 0.74 \\
    & \textbf{Ensembling}    & \textbf{0.84} & \textbf{0.86} & \textbf{0.96} & \textbf{0.65}& \textbf{0.30} &  \textbf{0.66} & \textbf{0.74} \\
    \hline
    \( \mathcal{M}_{(1,1)}^7 \) & Decision Tree & 0.84 & 0.84 & 0.95 & 0.80& 0.51 &  0.76 & 0.82 \\
    & Random forest & 0.84 & 0.85 & 0.96 & 0.81& 0.52 &  0.77 & 0.82 \\
    & XGBoost       & 0.86 & 0.88 & 0.94 & 0.75& 0.57 &  0.78 & 0.81 \\
    & LightGBM      & 0.86 & 0.88 & 0.94 & 0.75& 0.57 &  0.78 & 0.81 \\
    & \textbf{Ensembling}    & \textbf{0.86} & \textbf{0.88} & \textbf{0.95} & \textbf{0.75}& \textbf{0.57} &  \textbf{0.78} & \textbf{0.81} \\
    \hline
    \( \mathcal{M}_{(1,1)}^8 \) & Decision Tree & 0.83 & 0.87 & 0.93 & 0.60& 0.46 &  0.71 & 0.73 \\
    & \textbf{Random forest} & \textbf{0.82} & \textbf{0.84} & \textbf{0.95} & \textbf{0.73}& \textbf{0.44} &  \textbf{0.72 }& \textbf{0.78} \\
    & XGBoost       & 0.84 & 0.87 & 0.93 & 0.64& 0.47 &  0.72 & 0.75 \\
    & LightGBM      & 0.84 & 0.88 & 0.93 & 0.63& 0.47 &  0.72 & 0.75 \\
    & Ensembling    & 0.84 & 0.87 & 0.94 & 0.66& 0.47 &  0.73 & 0.75 \\
    \hline
    \( \mathcal{M}_{(1,1)}^9 \) & Decision Tree & 0.91 & 0.96 & 0.94 & 0.42& 0.56 &  0.72 & 0.69 \\
    & \textbf{Random forest} & \textbf{0.91} & \textbf{0.94} & \textbf{0.95} & \textbf{0.57}& \textbf{0.53} &  \textbf{0.75} & \textbf{0.75} \\
    & XGBoost       & 0.90 & 0.95 & 0.94 & 0.48& 0.50 &  0.72 & 0.71 \\
    & LightGBM      & 0.91 & 0.96 & 0.94 & 0.45& 0.53 &  0.72 & 0.70 \\
    & Ensembling    & 0.91 & 0.96 & 0.94 & 0.49& 0.58 &  0.74 & 0.70 \\

    \hline
  \end{tabular}$}
\end{table*}


In this section, we will showcase the performance of our models, leveraging one year's worth of data (i=1) to predict student outcomes in the following year (j=1). Our study primarily aimed at investigating the levels characterized by a high dropout rate. If the available data were sufficient to train the models, we decided to focus our analysis only on those levels for which we had a sufficient amount of data, focusing on building models for $k \in \{5, \ldots, 9\}$. The performance of our models using different splitting approaches is presented in Tables \ref{tab:results_random},\ref{tab:results_school}, and \ref{tab:results_years}.

Table \ref{tab:results_random} shows the performance of the models using the guided random split approach, over the different levels (k), denoted as $\mathcal{M}_{(1,1)}^k$, we observe that certain models outperform others. In particular, the XGBoost and LightGBM algorithms consistently show strong performance in terms of accuracy, recall, precision, F1-score, and AUC. Furthermore, as we move from one level to another, there is an improvement in the recall of the dropout class, i.e. the ability of the model to identify students at risk of dropping out. This improvement is in line with the differences in dropout rates observed between the different levels.

Using the split-by-school approach, as shown in Table \ref{tab:results_school}, we observe that there are a few differences, however, we find that the performance of our models closely matches the results obtained by the guided random split at all levels, underlining the robustness and stability of our models.

When using the split-by-year approach, as shown in Table \ref{tab:results_years}, we observe that the performance of the models is significantly lower compared to other approaches. This decrease in performance can be attributed to the influence of the COVID-19 pandemic, which brought about significant changes in the educational environment. As a result, both the dropout rates and the reasons for dropping out have shifted compared to typical years.

\subsection{Test on a class of student}
\begin{figure}[htb!]
  \centering
     \includegraphics[width=0.45\textwidth]{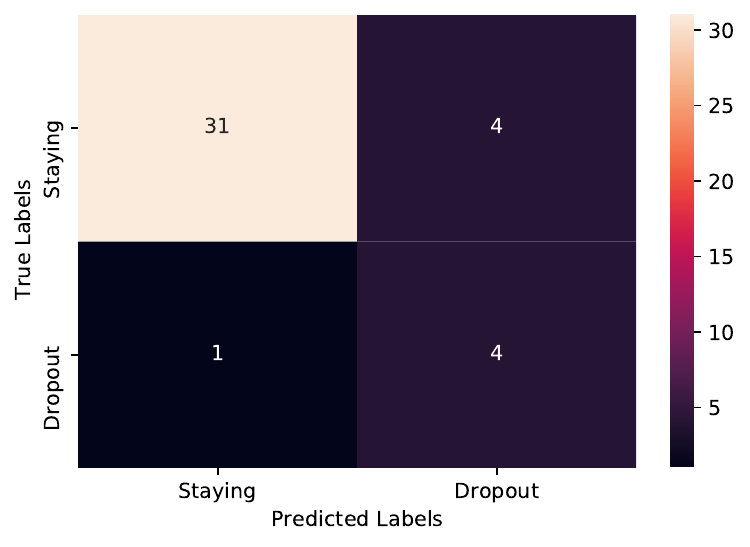}
  \caption{Confusion matrix for $\mathcal{M}_{(1,1)}^7$ in a class of 40 students }
  \label{fig:confusion_matrix}
  \end{figure}

Once deployed, the application will perform predictive analyses on a class of students. With this in mind, we have chosen to evaluate the impact of our models on a defined cohort of students to determine the statistical significance of the results. As shown in Figure \ref{fig:confusion_matrix}, the confusion matrix illustrates the performance of the \gls{LightGBM} model in the context of a class of 40 students, of which 5 students experienced attrition.

The confusion matrix shows that the model successfully identifies 4 out of 5 dropouts. However, we observe that the model incorrectly classifies 4 students who are not dropouts as potential dropouts. While our models have demonstrated their ability to identify dropouts, we acknowledge the need for additional support from experts in the field of education to address this issue effectively. As a proactive step, we present the predictions for each student along with their corresponding probability score of dropping out, as shown in Table \ref{tab:probability}.

\begin{table}[htb!]
    \centering
    \caption{True labels, predicted labels, and probability of dropout for student predicted as a dropout}
    \label{tab:probability}
    \begin{tabular}{ccc}
    \toprule
    True Label & Predicted Label & Probability of Dropout \\
    \midrule
    1.0 & 1.0 & 0.96 \\
    \textbf{0.0} & \textbf{1.0} & \textbf{0.62} \\
    \textbf{0.0} &\textbf{1.0} & \textbf{0.65} \\
    1.0 & 1.0 & 0.87 \\
    1.0 & 1.0 & 0.91 \\
    \textbf{0.0} & 1.0 & \textbf{0.77} \\
    1.0 & 1.0 & 0.65 \\
    \textbf{0.0} & 1.0 & \textbf{0.78} \\
    \bottomrule
    \end{tabular}
\end{table}

The fact that false predictions have a lower probability of dropping out than true dropouts suggests that the model tends to be cautious when making positive predictions. In other words, it is more likely to predict a student as a potential dropout if it is more confident (higher probability) in that prediction.  This reflects the model's attempt to avoid making confident predictions without strong evidence of dropout risk. However, it is important to strike a balance between recall and precision to ensure that true dropouts are not missed.

\subsection{The Impact of Prediction Corrector}

\begin{figure*}[htb!]
    \centering
    \includegraphics[width=0.8\textwidth]{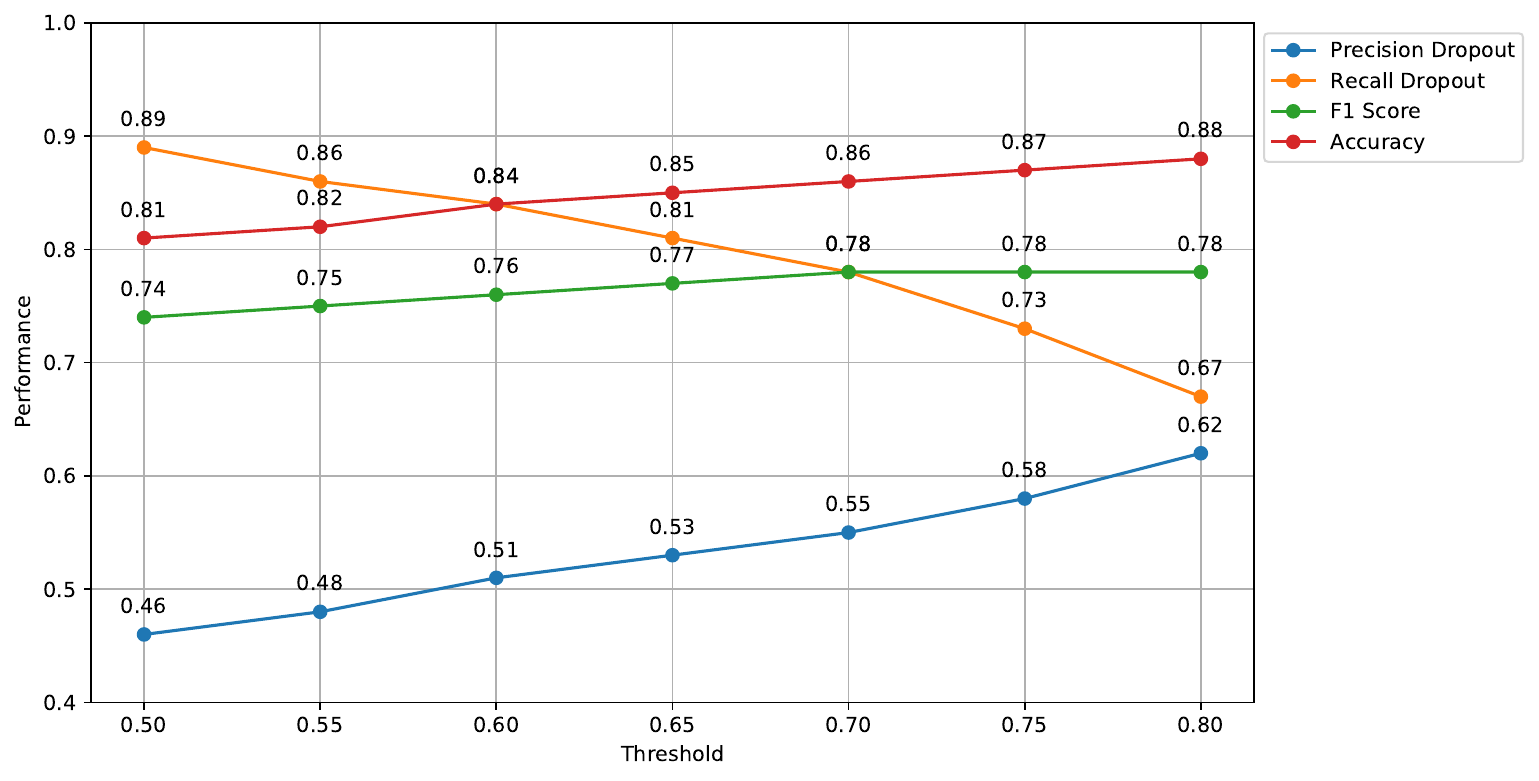}
    \caption{Impact of classification threshold on LightGBM $\mathcal{}_{(1,1)}^7$Performance }
    \label{fig:impact_of_threshold}
\end{figure*}

From Figure \ref{fig:impact_of_threshold} and Table \ref{tab:multiple_threshold}, we observe several interesting trends in model performance across different thresholds. As the threshold increases, both the accuracy and the F1-score of the model tend to increase. This indicates that the model's ability to correctly classify dropout and non-dropout instances improves as we increase the threshold. The precision of dropout predictions increases as the threshold increases. This means that when the model predicts a student as a dropout at higher thresholds, it is more likely to be correct in its prediction. On the contrary, the recall of dropout predictions tends to decrease as the threshold increases. This suggests that at higher thresholds, the model may miss some actual dropout cases, leading to a decrease in recall. In summary, adjusting the threshold value has a significant impact on the trade-off between precision and recall in dropout prediction. Higher thresholds result in more conservative dropout predictions with higher precision but lower recall, while lower thresholds lead to more inclusive predictions with higher recall but lower precision.

  \begin{table*}[htb!]
  \centering
  \caption{Performance of LightGBM $\mathcal{}_{(1,1)}^7$ using prediction corrector with multiple threshold classification}
  \label{tab:multiple_threshold}
  \begin{tabular}{lcccccccc}
    \toprule
      Threshold & Accuracy & \multicolumn{2}{c}{Class 0 (Continue)} & \multicolumn{2}{c}{Class 1 (Dropout)} &  F1-score & AUC \\
    \cmidrule(lr){3-4} \cmidrule(lr){5-6} 
     & & Recall & Precision  & Recall & Precision  &  \\
    \midrule
    0.50 & 0.81 & 0.80 & 0.97 & 0.89 & 0.46 &  0.74 & 0.83 \\
    0.55 & 0.82 & 0.82 & 0.97 & 0.86 & 0.48 &  0.75 & 0.83 \\
    0.60 & 0.84 & 0.84 & 0.96 & 0.84 & 0.51 &  0.76 & 0.83 \\
    \textbf{0.65} & \textbf{0.85} & \textbf{0.86} & \textbf{0.96} & \textbf{0.81} & \textbf{0.53} &  \textbf{0.77} & \textbf{0.83} \\
    0.70 & 0.86 & 0.88 & 0.95 & 0.78 & 0.55 &  0.78 & 0.82 \\
    0.75 & 0.87 & 0.90 & 0.94 & 0.73 & 0.58 &  0.78 & 0.81 \\
    0.80 & 0.88 & 0.92 & 0.93 & 0.67 & 0.62 &  0.78 & 0.79 \\
    \bottomrule
  \end{tabular}
\end{table*}

\subsection{Impact of Prediction Horizon (j) on Model Performance}

\begin{figure*}[htb!]
  \centering
  \includegraphics[width=1\textwidth]{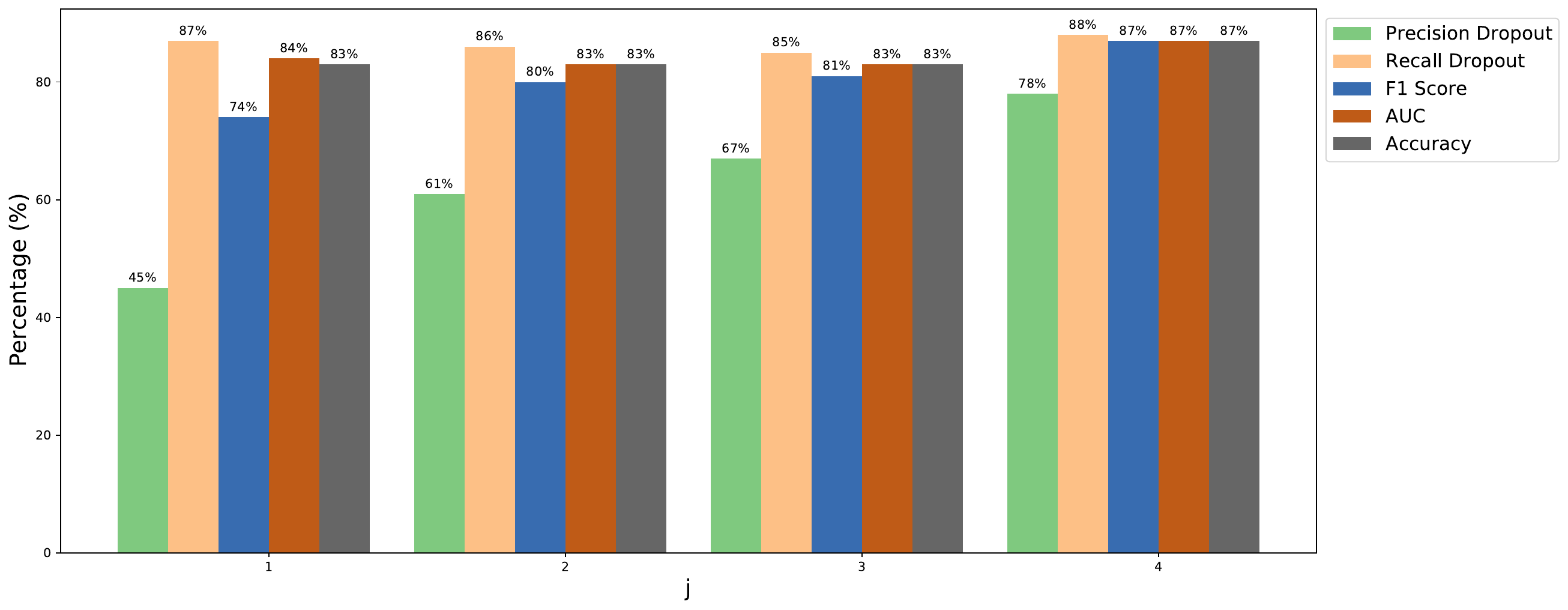}
  \caption{Impact of choosing prediction horizon (j) on $LightGBM\mathcal{}_{(1,j)}^7$ performance}  
  \label{fig:impact_of_j}
  \end{figure*}

In this section, we examine a crucial aspect of the proposed framework, the prediction horizon (j), which is the number of subsequent years for which the student's status is predicted, identifying the student's status in the coming (j) years. We therefore examine the impact of different prediction horizons on the overall performance of the models. Figure \ref{fig:impact_of_j} provides a comprehensive analysis of how the prediction horizon, affects the performance of machine learning models, with a primary focus on precision and recall for dropout prediction. A striking observation is the remarkable and consistent enhancement in the precision of identifying dropout students as the prediction horizon (j) increases. This suggests that a longer prediction horizon allows the model to capture more complex dropout patterns and behaviours.

Furthermore, the F1 score, which harmonizes precision and recall, shows a robust positive correlation with (j). As (j) progressively increases, the F1 score consistently increases. This phenomenon implies that the model's ability to provide accurate predictions for both dropouts and non-dropout students becomes increasingly balanced and effective as the prediction horizon increases.

\subsection{Impact of Historical Data Duration (i) on Model Performance}

In this section, we explore another key element of our proposed framework, the number of years (i) used, and its impact on the model's performance. The number of years used represents the temporal range of historical data that forms the basis of our model's predictive capabilities. By investigating the effect of the number of years used on the performance of our model, we aim to identify the optimal time frame that produces the most accurate and reliable results.

  \begin{figure*}[htb!]
  \centering
  \includegraphics[width=0.9\textwidth]{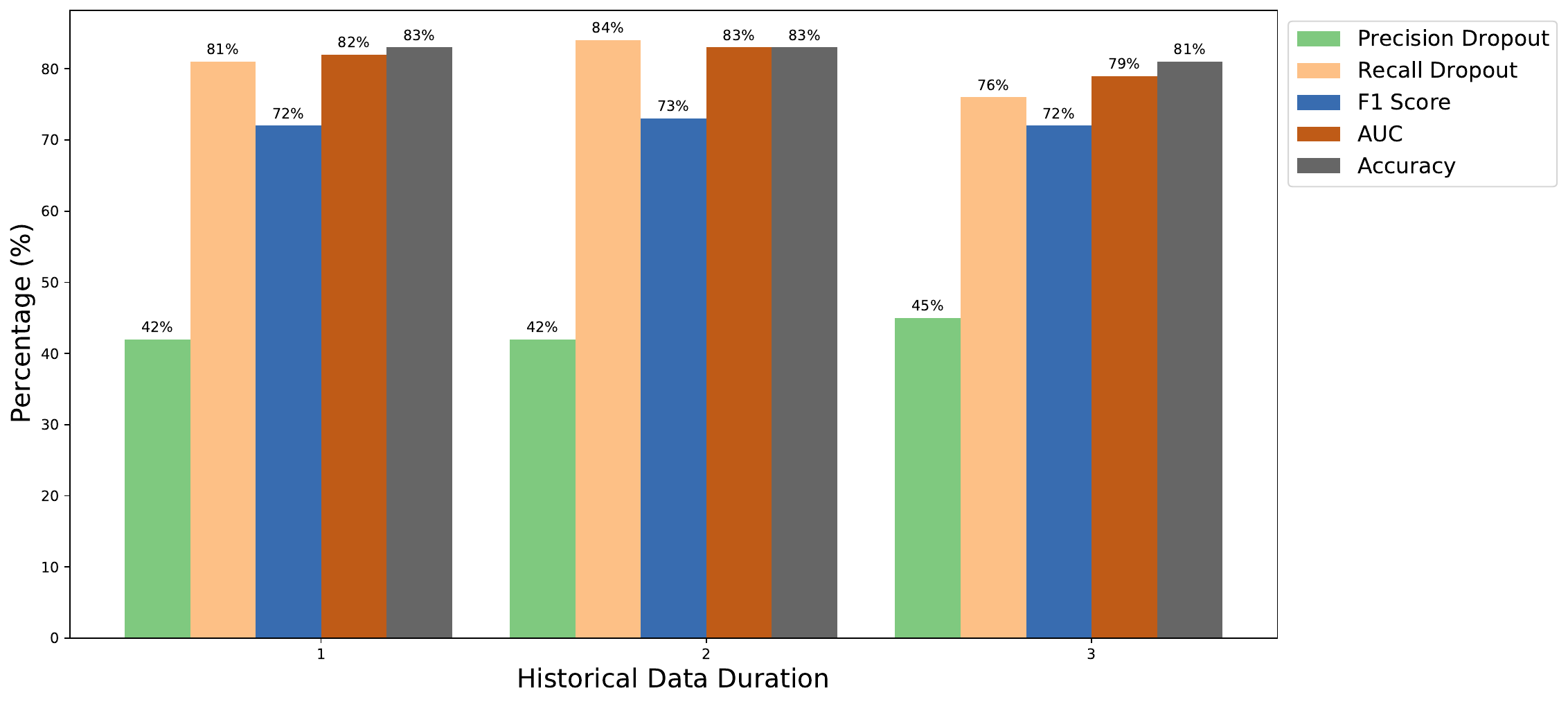}
  \caption{Impact of choosing historical data duration (i) on $LightGBM\mathcal{}_{(i,1)}^9$ performance } 
  \label{fig:impact_of_i}
  \end{figure*}

In this study, we explored the option of aggregating the dataset (aggregating the values of the same feature) as an alternative to concatenation. However, after rigorous analysis and experimentation, we found that both methods gave similar performance results. Ultimately, we decided to use concatenation to preserve the information from each year, as this approach allows for a more comprehensive interpretation of the results. When we want to use more than one year of historical data, we concatenate the data from all the years under consideration.  However, it is possible that the data from the same year before the target year may not relate to the same academic level, as some students may have experienced academic setbacks and not progressed to the intended level. To address this potential issue, we use the feature that indicates the academic level for each year of historical data included in our analysis.

As shown in Figure \ref{fig:impact_of_i}, we can observe the impact of the number of years of historical data on various performance metrics for a dropout prediction model. Specifically, we examine the impact on precision, recall, F1-score, and AUC, while also considering accuracy. The precision of the dropout class tends to increase as we use more years of historical data (i-values). Interestingly, when we use 2-years of historical data (i=2), we notice a significant increase in both recall and the F1-score for the dropout class. This suggests that a 2-year historical dataset provides a balance between recall and maintaining a good overall F1-score. Similarly, the area under the ROC curve (AUC) also increases when using 2-years of historical data. This suggests that 2-years of data results in better model discrimination between dropout and non-dropout cases. However, when we use 3-years of historical data (i=3), we observe a decrease in the performance of the model. Precision, recall, F1-score, and AUC decrease, suggesting that an excessive amount of historical data may lead to overfitting or introducing noise into the model, resulting in poorer predictive performance.

In summary, the choice of the number of years of historical data significantly impacts the performance of the dropout prediction model. While using more data generally improves precision, a balanced choice, such as 2-years of historical data, tends to provide better overall performance in terms of recall, F1-score, and AUC. However, using too much data, as in the case of 3 years, can lead to decreased model performance.

\subsection{Interpretation of the model's predictions}
\begin{figure*}[htb!]
  \centering
  \includegraphics[width=0.9\textwidth]{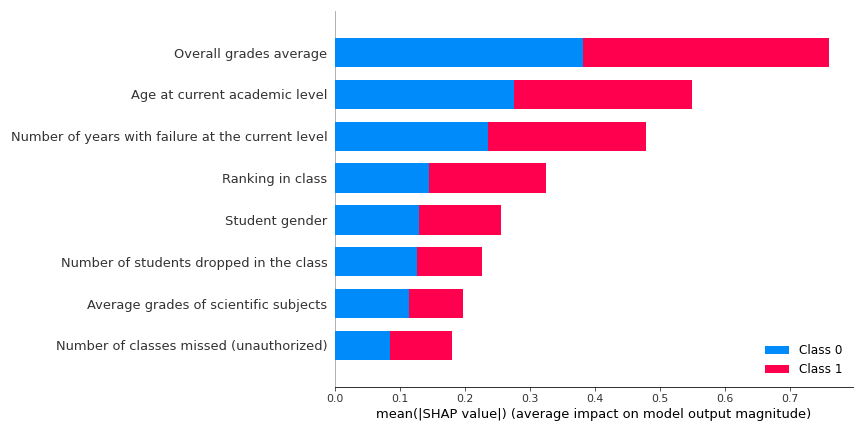}
  \caption{Top 8 important features according to the \gls{SHAP} technique}
  \label{fig:Shape}
\end{figure*}

Using the \gls{XAI} techniques allows us to identify the key factors influencing the model's output, as shown in Figure \ref{fig:Shape}. In particular, our analysis using \gls{SHAP} explanations shows that features such as overall grade average, age, gender, ranking in the class, and other characteristics emerge as prominent contributors. This insight provides a comprehensive understanding of the influence of each characteristic on the model's predictions.
Leveraging this knowledge, in conjunction with domain expertise in the field of education, will enable us to develop robust support strategies. By identifying and understanding the specific impact of these characteristics, we can tailor our educational interventions and initiatives more effectively, ultimately improving the quality of support provided.

\section{Discussion}\label{sec:Discussion}

In the realm of addressing the pervasive problem of student dropout, numerous solutions and methodologies have been proposed in existing literature. While these efforts have undoubtedly shed light on various facets of the issues, such as imbalanced data problems, and limited evaluation strategies, It is worth noting that a universal and all-encompassing solution is difficult because the landscape of education systems is inherently complex and dynamic, and dropout is influenced by a variety of factors, both individual, family, and academic. Our research provides a unique perspective by introducing a predictive modeling approach that can be applied to many education systems, and the proposed evaluation strategy and comparison of imbalanced data techniques can help future works to be more concerned about these issues.
Despite the promising results of our research, it is important to acknowledge the limitations of our study. One notable limitation relates to the precision of predicting the exact level or year of dropout. While our models excel at identifying students at risk, pinpointing the exact moment of dropout remains a challenging task. The educational journey is influenced by several unpredictable circumstances, and students may drop out at different stages. Future research could explore methods to improve the temporal accuracy of dropout predictions.

To further advance the field of dropout prediction and provide even more targeted support to at-risk students, future research should delve into survival analysis techniques. Survival analysis is a specialized statistical method that accounts for the time-to-event data, making it particularly apt for predicting the timing of student dropouts accurately. By incorporating survival analysis into predictive modeling, researchers can gain deeper insights into the temporal aspects of dropout and identify critical intervention points. This would enable educational institutions to implement timely measures precisely when they are needed most, maximizing their impact on student retention. In addition, the recommendation system has been used in education \cite{urdaneta2021recommendation,mpia2023cobert} and various other areas \cite{MEKOUAR20221}, \cite{l4806899l}. We will work on building a recommendation system that can help students with effective recommendations based on student profiles \cite{mekouar2023global}.

In summary, this study presents a tailored predictive modeling approach that can be used by many education systems around the world. There is a need for future research to explore advanced methods, such as survival analysis, to refine and enhance dropout prediction accuracy. By collectively building upon these research efforts, we can work toward a future where educational institutions possess the tools and insights needed to proactively support students at risk of dropout, ultimately fostering improved educational outcomes for all.

\section{Conclusion}\label{sec:Conclusion}
In conclusion, this research addresses a pressing global issue, student dropout, which varies significantly across countries due to a variety of academic, socio-economic, and family factors. This paper presents a pioneering predictive modeling approach tailored to identifying students at risk of dropping out within education systems. Using an extensive dataset provided by the Moroccan Ministry of National Education, Preschool and Sports, our method integrates a wide range of demographic, academic, and institutional characteristics. These features, meticulously extracted from the ministry's comprehensive data management system, allow us to develop a robust and versatile solution using state-of-the-art machine learning techniques. The ultimate goal of our research is to accelerate timely interventions and support mechanisms for students at risk of dropping out. By identifying these students early in their educational journey, we aim to increase student retention rates and ultimately improve educational outcomes across the educational landscape. Notably, the methodology we propose is remarkably versatile, making it applicable across different education systems and at all levels of study. The evaluation strategy we have introduced provides invaluable insight into the performance of our models, particularly their effectiveness in predicting dropout cases. Our most robust model achieves impressive metrics, including an accuracy rate of 88\%, a recall of 88\%, a precision of 86\%, and an AUC (area under the curve) of 87\%. These results underline the potential utility and effectiveness of the method in a variety of educational contexts. In essence, our research not only contributes to the broader discourse on tackling student dropout but also presents a compelling case for the applicability of our methodology in diverse educational settings. By harnessing the power of data and machine learning, we are taking significant steps towards achieving more equitable and successful educational outcomes for students, both in Morocco and beyond.

\bibliographystyle{unsrt}  
\bibliography{references}

\end{document}